\documentclass{aa}
 \usepackage{amssymb}

\usepackage[colorlinks, citecolor=blue]{hyperref}
\usepackage{multirow}
\usepackage{rotating}
\usepackage{ctable}
\usepackage{array}
\usepackage{longtable}
\usepackage{tabularx}
\usepackage{rotating}

\usepackage{graphicx,epsfig,fancyhdr,psfig,rotating,amsmath,epsf,txfonts,natbib,epstopdf,multirow}
\begin{document}

\title{High--cadence observations of spicular-type events on the Sun}

\author{J. Shetye$^1$, J.G. Doyle$^1$, E. Scullion$^{2,7}$, C. J. Nelson$^{1,3,4}$, D. Kuridze$^4$, V. Henriques$^4$, F. Woeger$^5$, T. Ray$^6$}

\offprints{jus@arm.ac.uk}
\institute{$^1$Armagh Observatory, College Hill, Armagh BT61 9DG, N. Ireland \\
$^2$School of Physics, Trinity College Dublin, Dublin 2, Ireland\\
$^3$Solar Physics and Space Plasma Research Centre, University of Sheffield, Hicks Building, Hounsfield Road, Sheffield, S3 7RH, UK\\
$^4$Astrophysics Research Centre, School of Mathematics and Physics, Queen's University Belfast, BT7~1NN, N. Ireland\\
$^5$NSO/DKIST, University of Colorado at Boulder, 3665 Discovery Drive, Boulder, CO 80303\\
$^6$ Dublin Institute for Advanced Studies, 31 Fitzwilliam Place, Dublin 2, Ireland \\
$^7$ Department of Mathematics \& Information Sciences, Northumbria University, Newcastle Upon Tyne, NE1 8ST, UK}
 \date{Received date, accepted date}

\abstract
{Chromospheric observations taken at high-cadence and high-spatial resolution show a range of spicule--like features, including Type--I, Type--II (as well as RBEs and RREs which are thought to be on--disk counterparts of Type--II spicules) and those which seem to appear within a few seconds, which if interpreted as flows would imply mass flow velocities in excess of 1000 km s$^{-1}$.}
{This article seeks to quantify and study rapidly appearing spicular-type events. We also compare the MOMFBD and speckle reconstruction techniques in order to understand if such spicules are more favourably observed using a particular technique.}
{We use spectral imaging observations taken with the CRisp Imaging SpectroPolarimeter (CRISP)  on the Swedish 1--m Solar Telescope. Data was sampled at multiple positions within the H$\alpha$ line profile for both an on-disk and limb location.}
{The data is host to numerous rapidly appearing features which are observed at different locations within the H$\alpha$ line profile. The feature's durations vary between 10 -- 20 s and lengths around 3500 km. Sometimes, a time delay in their appearance between the blue and red wings of 3 -- 5 s is evident, whereas, sometimes they are near simultaneous. In some instances features are observed to fade and then re--emerge at the same location several tens of seconds later.}
{We provide the first statistical analysis of these spicules and suggest that these observations can be interpreted as the line--of--sight (LOS) movement of highly dynamic spicules moving in and out of the narrow 60 m\AA\ transmission filter used to
observe in different parts of the H$\alpha$ line profile. The LOS velocity component of the observed fast chromospheric features, manifested as Doppler shifts, are responsible for their appearance in the red and blue wings of H$\alpha$  line.
Additional work involving data at other wavelengths is required to investigate the nature of their possible wave--like activity.}

\authorrunning{Shetye, J. et al.}
\titlerunning{High cadence spicular events.}

\maketitle

\section{Introduction} 
\label{intro}
\begin{figure}
\centering
\includegraphics[scale=0.30]{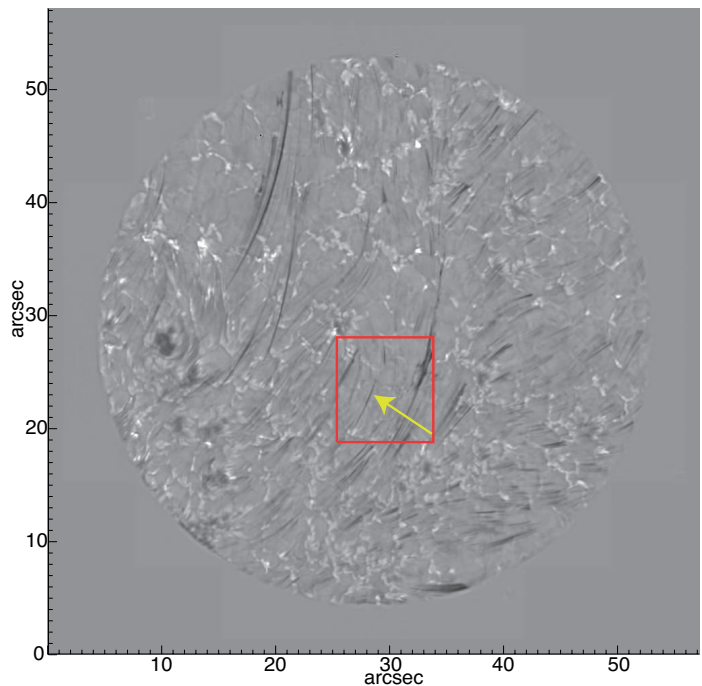}

\caption{An image reconstructed using the speckle technique (the CRISP FOV is changed from square to circle to allow better reconstruction for speckling) from the 10 June 2014 dataset. The section framed in red is shown in more detail in Fig.~\ref{comp}. The arrow indicates the location of a feature of interest. The CRISP FOV should be tilted at an angle of +62$^{\circ}$04$\arcmin$.} \label{sROI}
\end{figure}

Spicules are highly dynamic jet-like structures best observed in the chromospheric absorption lines such as H$\alpha$ (6563 \AA), CaII H (3968.5 \AA) and Ca II (8542 \AA). These features exhibit apparent lengths between 0.5 Mm and 15 Mm and could potentially contribute to atmospheric mass cycling and heating of the corona. Type I spicules have lifetimes between 3 and 7 mins and exhibit up and downward motions with speeds of 20 -- 50 km s$^{-1}$. \citet{2007PASJ...59S.655D} noted the existence of relatively faster ($\approx$100 km s$^{-1}$) and shorter lifetime (50 s) spicules, which they called Type--II spicules, however \citet{2012ApJ...750...16Z} questioned whether spicules should be divided into two classes. A comprehensive study by
\citet{2012ApJ...759...18P} using quiet Sun, coronal hole and active region datasets quantifies various properties of spicules revealing a bi-modal distribution of key properties suggesting that there may indeed be two varieties of spicules. In a follow up study \citet{2013ApJ...764...69P}, stated that Type--II could have been missed due to the low temporal and spatial resolutions of previous instruments.

Type--I spicules are thought to result from photospheric $\it{p}$-mode leakage between granular cells, hence supplying energy and power \citep[see reviews by][and references there in]{2009SSRv..149..229T} into the mid-chromosphere, across the temperature minimum (4000\,K) region. Material is then thought to propagate along magnetic flux tubes which are inclined to the vertical (typically at an angle of 40 -- 50$^o$), e.g. see \citet{2004Natur.430..536D,2007Sci...318.1572E}. The acoustic $\it{p}$--mode oscillation periods would normally be expected to be evanescent in the chromosphere, however, gravity along the inclined field of a magnetic wave guide increases the acoustic cutoff period, subsequently allowing $\it{p}$--modes to tunnel through and penetrate directly into the hotter upper chromosphere \citep{2004Natur.430..536D}.

Type--II spicules could be formed by magnetic reconnection \citep{2008ApJ...679L.167L,2007Sci...318.1574D}. \citet{2007ApJ...655..624D} suggested a Type--II spicule can reach coronal temperatures, whereas \citet{2011A&A...532L...1M} showed that
even the fastest spicules did not form at temperatures higher than \textbf{$3\times10^6$} K. At present, the contribution of Type--II spicules to coronal heating is still under debate \citet{2012JGRA..11712102K,2013ApJ...779....1T,2014ApJ...781...58P}. However, it is interesting to note that \citet{2014ApJ...792L..15P} combined datasets from Interface Region Imaging Spectrograph \citep[IRIS]{2014SoPh..tmp...25D}, the Hinode Solar Optical Telescope (SOT) and the Solar Dynamics Observatory's Atmospheric Imaging Assembly \citep[AIA]{2012SoPh..275...17L} to show that Type--II spicules undergo thermal evolution at least to the transition region. Recent work by \citet{2015ApJ...799L...3R} explored this problem using a combination of IRIS data and ground-based data from the Swedish Solar Telescope \citep[SST]{2003SPIE.4853..341S}, linking features found in the chromosphere to the transition region. Once again this provides a means of linking spicules across multiple atmospheric levels. 

Rapid Blue--shifted Excursions (RBEs) were first reported by \citet{2008ApJ...679L.167L}, while trying to find on-disk counterparts of the Type--II spicules. Additional studies have also been carried out by, for example, \citet{2009ApJ...705..272R}. RBEs are short lived events (lasting $\approx$ 45 s) observed as sudden broadenings in the blue-wing of chromospheric spectral lines. Their average length is 3.0 Mm with widths of 0.250 Mm and Doppler velocities between 15 -- 20 km s$^{-1}$ as was reported by \citet{2013ApJ...764..164S,2013ApJ...769...44S}. Those authors reported the presence of absorption features in the red wing of the Ca II 8542 \AA\ and H$\alpha$ lines, referring to them as rapid redshifted excursions (RREs). Further, they suggested that there are three kinds of motion in the chromospheric flux tubes, i.e. field--aligned flows, swaying motions and torsional motions. \citet{2015ApJ...802...26K} showed that RREs and RBEs have similar occurrence rates, lifetimes, lengths, and widths. These features showed non-periodic, nonlinear transverse motions perpendicular to their axes at speeds of 4 -- 31 km s$^{-1}$. Furthermore, both types of structures either appeared as high speed jets that are directed outward with speeds of
50 -- 150 km s$^{-1}$, or emerged within a few seconds. 
 
\citet{2011ApJ...730L...4J} reported the appearance of spicule--like features which only lasted a few seconds, implying un--realistic flows of around 1000 km s$^{-1}$ in the chromosphere. This was followed by additional work by \citet{2012ApJ...755L..11J}, resulting in the idea that some spicules may be formed in a sheet--like structures of
chromospheric material and therefore their appearance, or disappearance, within a few seconds was simply due to a
line-of-sight effect. \citet{2014ApJ...785..109L} reported similar structures in the H$\alpha$ blue and red wings near the solar limb. These structures occurred with comparable occurrence rates without any evidence for torsional motions. They presented arguments that at least some of the detected structures corresponded to warps in two--dimensional sheets. 

In this article, we investigate these apparent ultra-fast spicules using spectral imaging data from the CRisp Imaging SpectroPolarimeter (CRISP) on the Swedish Solar Telescope  (SST) for both disk and limb observations including a specially designed observing sequence to test whether what is seen could be due to the data reduction method employed.  In Sect.~\ref{obs} we describe the observations while the data reduction and code comparison are given in Sect.~\ref{dR} and Sect.~\ref{DaCC} respectively. The data analysis and statistics are given in Sect.~\ref{DnS}, while our conclusions are given in Sect.~\ref{Diss}. Finally, a summary is presented in Sect.~\ref{Summ}.

\section{Observations}\label{obs}

\begin{figure*}
\centering
\includegraphics[scale=0.40]{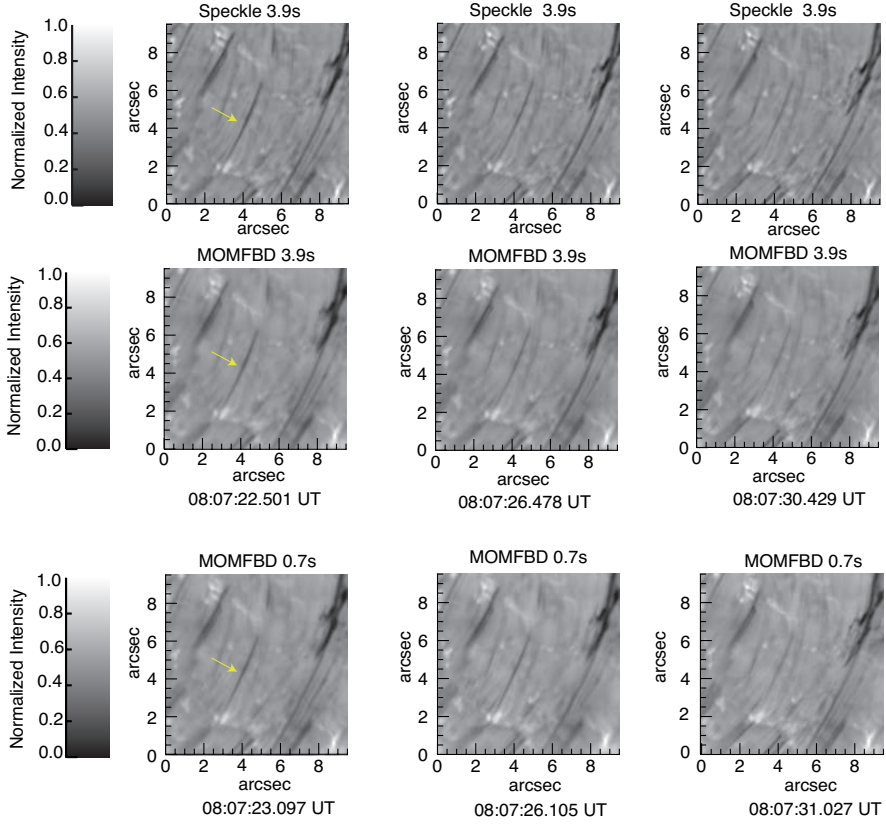}

\caption{Three normalized consecutive images (horizontal direction) constructed using three different data reduction techniques  (see Sect.  \ref{dR}) for the 10 June dataset taken in the H$\alpha$ blue wing at --774 m\AA. The speckled data and MOMFBD (ML) data is constructed at 3.9 s cadence whereas the lower panel's MOMFBD (MH) data is constructed at 0.7 s cadence. Speckled and MOMFBD 3.9 s panels are taken at the same time as indicated at the bottom of the middle panel. The MOMFBD 0.7 s images are taken at the time closest to the time of the images in the rows above. The yellow arrows points towards the location of a feature of interest}\label{comp}
\end{figure*}

\begin{figure*}
\centering
\includegraphics[scale=0.70]{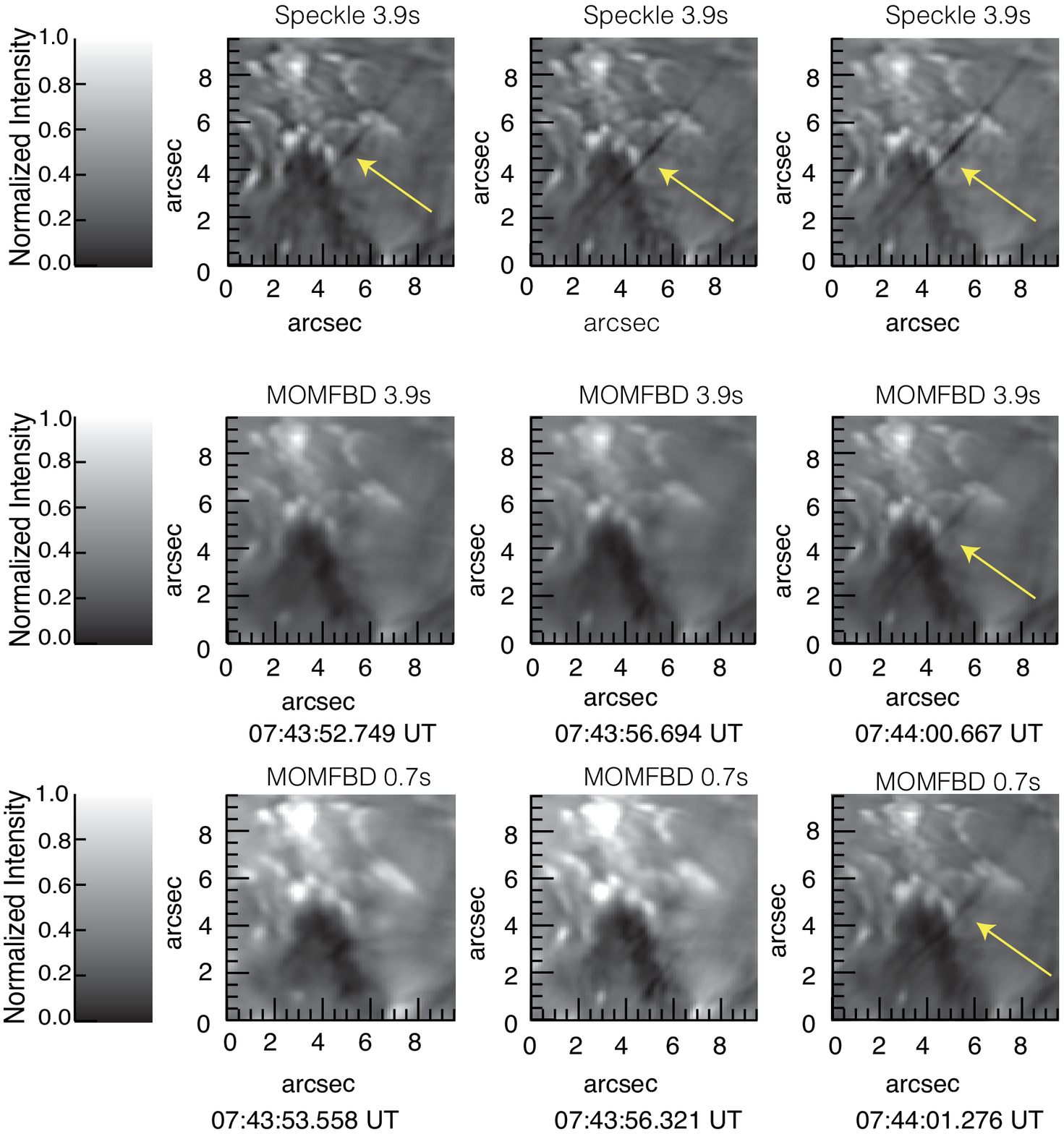}

\caption{Three normalized consecutive images constructed (horizontal direction) using the three different data reduction techniques  (see Sect. \ref{dR}) for the 10 June data taken in the H$\alpha$ blue wing at --0.774m\AA. Here we see that a spicule is detected by all three techniques, . The yellow arrows indicate the feature of interest. }\label{compmissing}
\end{figure*}

We present spectral imaging observations taken using the CRISP \citep{2008ApJ...689L..69S} instrument on the Swedish 1--m Solar Telescope  \citep[SST]{2003SPIE.4853..341S}. CRISP has a field of view (FOV) of 60$\arcsec$ $\times$ 60$\arcsec$ with an approximate pixel scale of 0.0592$\arcsec$. After co-alignment with SDO/AIA 170.0 nm images, we determined that
the CRISP FOV is tilted at an angle of \textbf{62$^{\circ}$04$\arcmin$} with respect to the SDO heliocentric FOV. The diffraction limited resolution at H$\alpha$ is 0.14$\arcsec$. The H$\alpha$ pre--filter FWHM is 490 m\AA\, while the FWHM of the transmission filter is 60.4 m\AA\ allowing observations at multiplet positions within the H$\alpha$ line profile.  The telescope uses an adaptive optic system consisting of a tip--tilt mirror and a deformable mirror to correct for the effects of atmospheric seeing. After data acquisition the flats, darks and gains are used to produce quick look images free of fringe effects etc. Then we applied a pinhole calibration to correction for spatial offsets between the cameras (CRISP--T. CRISP--R and wideband (WB) camera). Data from all three cameras were then passed into an image reconstruction algorithm to correct for high frequency distortions before the images were destretched to the wideband (WB) as an anchor channel and then derotated to correct for solar rotation. Finally image scale corrections were applied to compare between different wavebands (such as H$\alpha$ versus \ion{Ca}{II} 8542 \AA\,) on a pixel--by--pixel basis. At the telescope the zero wavelength was calibrated at disk center. The "line--core" wavelength of the quiet sun "averaged background profile" is ascribed as "the rest wavelength" in calculating the Doppler velocity by comparison with profiles for various spectral pixels which evolve in time. In the following subsections, we discuss the two datasets analysed in this article.

\subsection{10 June 2014: datasets A \& B.} \label{data}

The first observations analysed in this article were collected during a 50 minute interval of good seeing between 07:17 UT and 08:08 UT on 10th June 2014 using the SST/CRISP instrument.  The approximately $60\arcsec\times60\arcsec$ FOV of these data was centered at $x_\mathrm{c}= 403\arcsec$ and $y_\mathrm{c}=-211\arcsec$ which contained a pore and was entirely within NOAA AR 12080. Nine H$\alpha$ line--positions were sampled in sequence, with eight images being collected at each of $\pm$ 1032, $\pm$ 774, $\pm$ 516, $\pm$ 258 m\AA\, (relative to the line core at $6562.8$ \AA\, at 0) before 36 frames were collected at each of $-774$ m\AA\, and the line core. These two positions were selected for additional sampling as $-774$ m\AA\, is the predominant spectral position for observing RBEs and the H$\alpha$ line core provides an excellent overview of the chromosphere. In the following sections, the nine point H$\alpha$ line scan will be denoted dataset A and the two positions sampled with $36$ frames will be referred to as dataset B. 

The number of frames used within dataset B is higher than the number used for dataset A due to the differences in reduction methods applied. Dataset A was reduced using the Multi-Object Multi--Frame Blind Deconvolution (MOMFBD) method \citep{2005SoPh..228..191V} which is capable of dealing with low numbers of frames whereas dataset B was reduced using the speckle method which required a statistically high number ($30-100$) of images taken using a high cadence \citep[and references therein]{2008A&A...488..375W}. The combination of these methods allows us to conduct a thorough analysis of events without introducing data reduction specific effects. The effective cadence for both dataset A and B was 3.9 s, however, it was also possible to produce four MOMFBD images ($8\rightarrow1$) for each repetition of dataset B giving a cadence of $0.7$ s (the highest achievable cadence) over these frames. These observational procedures were specifically designed for the detection of spicules, RBEs, RREs and the apparent ultra-fast spicules discussed in this article.
An example frame from the 10 June dataset constructed using the speckle technique is plotted in Fig.~\ref{sROI}. This image was sampled in the H$\alpha$ blue wing at -774 m\AA\,. A representative spicule feature discussed in this article is highlighted by the yellow arrow in the red box. Comparisons of results obtained using datasets A \& B are made in Fig.~\ref{comp} and Fig.~\ref{compmissing}.

\subsection{05 June 2014}

The second dataset analysed in this research was collected by the SST/CRISP instrument on 05 June 2014 between 11:53 UT and 12:34 UT. The $60\arcsec\times60\arcsec$ FOV sampled during this time was centred on $x_\mathrm{c}=876\arcsec$, $y_\mathrm{c}=343\arcsec$ meaning these data incorporated the solar limb. This sequence consisted of nine H$\alpha$ line positions at $\pm$ 1032, $\pm$ 774, $\pm$ 516, $\pm$ 258 m\AA\, and line centre at $6562.8$ \AA\,) and  nine Ca~{\sc ii} ($8542$ \AA\,) line positions (at $\pm$ 385, $\pm$ 165, $\pm$ 110, $\pm$ 55 m\AA\, and 0 m\AA\, corresponding to line centre). Only the H$\alpha$ line scans are analysed in the current work and, therefore, the Ca~{\sc ii} shall not be further discussed. These data were reduced in a manner similar to dataset A in the 10 June dataset meaning eight frames were taken at each of the above nine wavelength positions. After MOMFBD reconstruction, an effective science--ready cadence of 5 s was achieved meaning these data are perfect for analysing rapid dynamics in the chromosphere.

\section{Data reduction} \label{dR}

As mentioned above, we used two post-facto image restoration techniques: MOMFBD image restoration \citep{2006ASPC..354...55V} and the phase division speckle technique \citep{2008A&A...488..375W}, which are both discussed in more detail below. After processing, we applied a destretching algorithm to create the final images. With the MOMFBD method we followed the standard procedures in the reduction pipeline for CRISP data \citep{2015A&A...573A..40D}, which includes the post--MOMFBD correction for differential stretching as suggested by \citet{2012A&A...548A.114H}. We explored the fully processed datasets with CRISPEX \citep{2012ApJ...750...22V}, which is a versatile code for analysis of multi--dimensional spectral data-cubes and Timeslice ANAlysis tools (TANAT) for analysis and measurements of data. 

\begin{figure}
\centering
\includegraphics[scale=0.40,angle = 270]{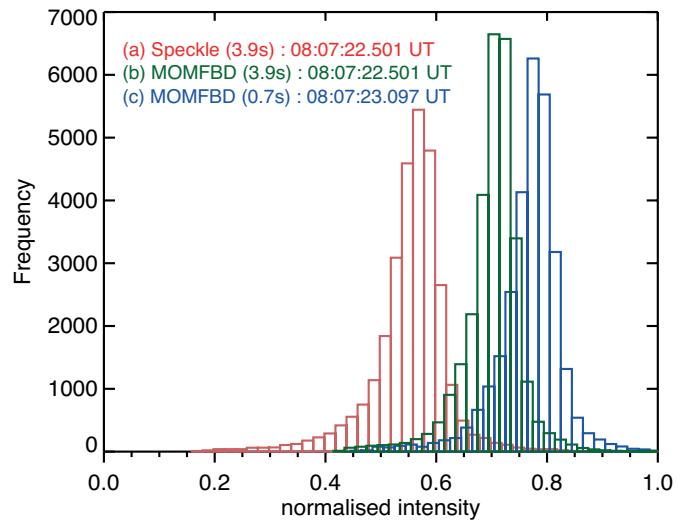}
\caption{Histogram plots for the three datasets from 10 June 2014 for the feature shown in Fig.~\ref{comp} showing the number of pixels as a function of the normalised intensity. The red histogram is made using speckle reconstruction, the dark--green histogram is made using MH and the blue histogram is made using ML.}\label{Hist}
\end{figure}
\begin{figure}[h!]
\centering
\includegraphics[scale=0.40,angle = 270]{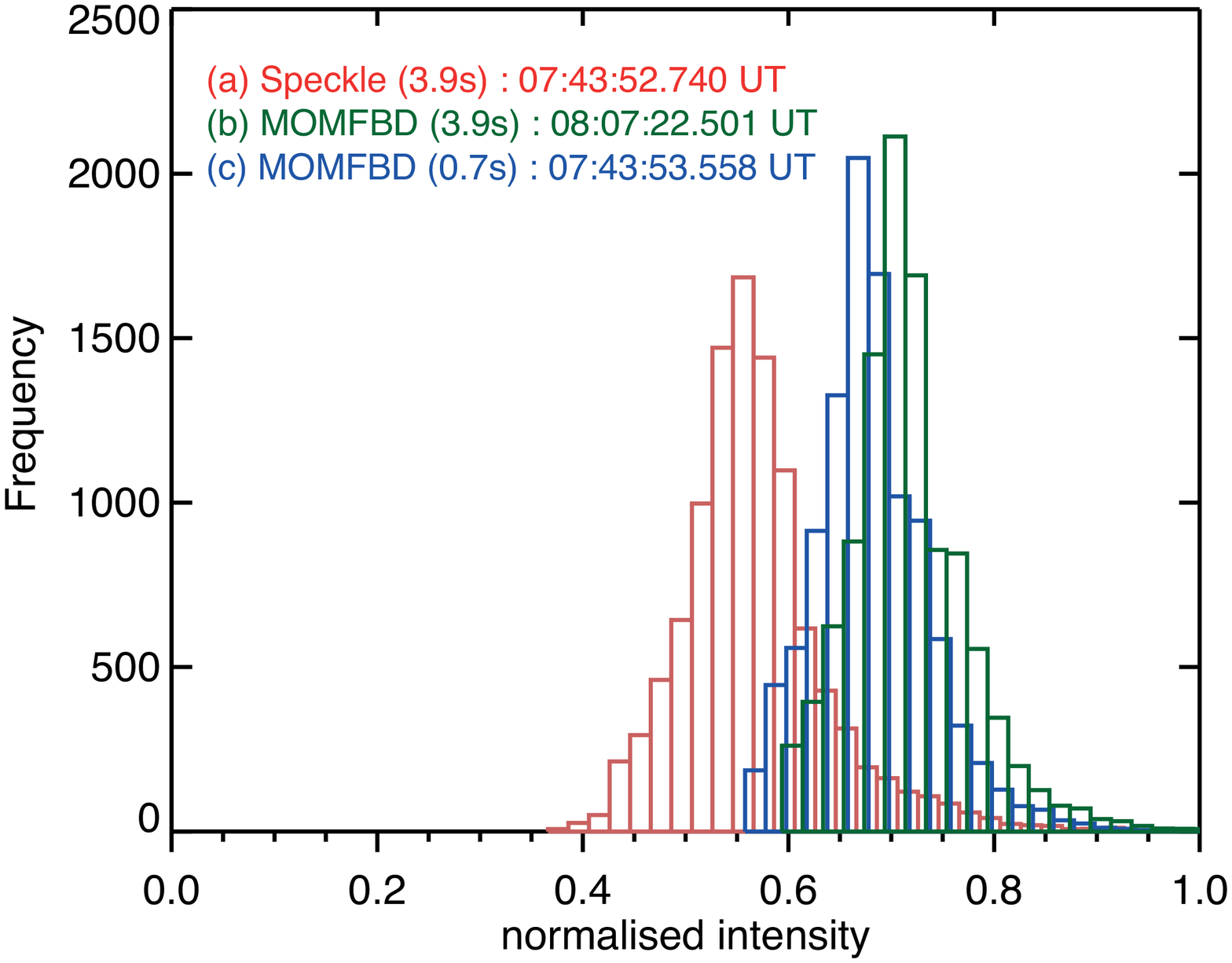}

\caption{Histogram plots for the three datasets from 10 June 2014 for the feature shown in Fig. \ref{compmissing} showing the number of pixels as a function of the normalised intensity. The red histogram is  made using speckled reconstruction, the dark--green histogram is  made using MH and the blue histogram is made using ML.}\label{Hist_comp}
\end{figure}

\begin{figure}[h!]
\centering
\includegraphics[scale=0.45]{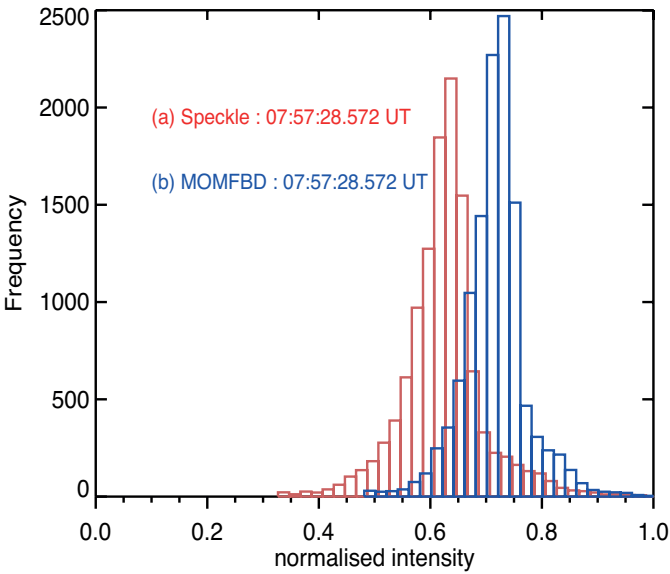}
\caption{Histogram plot for images constructed using speckle and MOMFBB with the same number of frames. The red histogram is  made using speckled reconstruction and the blue histogram is made using ML with 44 frames.}\label{Hist_comp_44}
\end{figure}

\subsection{Speckle reconstruction of the 10 June data} \label{Speckle}

Speckle interferometry was developed in the mid-1970s and was modified and improved further by various authors since then \citep{2008A&A...488..375W}.  This method uses a number of short-exposed images at a single wavelength position to compute an estimate of the unperturbed object image by evaluation of either the cross- or the bi-spectrum of the data. The observed images represent the image of the real object modulated by the instantaneous optical transfer function (OTF) of the Earth's atmosphere and the telescope \citep{2008A&A...488..375W}.

The reduction of speckle data in this paper is done using the Kiepenheuer--Institute Speckle Interferometry Package (KISIP) \citep{2008A&A...488..375W,2008SPIE.7019E..1EW}. This program uses Fourier phase reconstruction algorithms with either an extended Knox--Thompson or a triple correlation scheme. The images are separated into subframes based on an isoplanatic patch which enables a straight forward use of parallel processing pipeline. The package separates the image's fourier phases from its amplitudes and treats reconstruction of the object Fourier amplitudes separately. This involves steps such as, phase reconstruction based on an algorithm by \citet{1974ApJ...193L..45K} and amplitude reconstruction as developed by \citet{1970A&A.....6...85L}. 

In this section, we present the results of using the data sets described in Section.~\ref{data} in the following way: dataset A's 8 H$\alpha$ --774 m\AA\, images were combined with the 36 images of the subsequent dataset B's burst of H$\alpha$ --774 m\AA\, images, resulting in a burst of 44 images at H$\alpha$ --774 m\AA\,. Similarly, dataset A's H$\alpha$ line core images were combined with the 36 images of the subsequent dataset B's burst of H$\alpha$ line core images, resulting in a burst of 44 images at the H$\alpha$ line core wavelength. This processing was performed for all data acquired for the duration of the observation (50 minutes), resulting in many "image bursts" with an effective cadence of 3.9s. In a final step, these images bursts were reconstructed to single images using the triple correlation algorithm described in \citet{2008A&A...488..375W}, using a sub--field size of 5$\arcsec\times$5$\arcsec$ (128 $\times$ 128 pixels) and a spatial frequency cutoff ($''$phase reconstruction limit$''$) set to 75$\%$ of the diffraction limit. All other parameters were kept at their default values.

\subsection{Multi--object multiframe blind deconvolution (MOMFBD) for data1 and data2} \label{MOMFBD}

For both the 5 \& 10 June  datasets, we created MOMFBD images for all 9 H$\alpha$ 
line positions resulting in a final cadence of 3.9s (henceforth ML) and 5.0s respectively. Since we
have 36 images taken at both H$\alpha$ line centre and H$\alpha$ wing of  --774 m\AA\ in the 10 June data, we can create an additional 4 MOMFBD images resulting in an
effective cadence close to 0.7s (henceforth MH) for these five images followed by a 3.9s gap
before the next series of images were generated. 
   
\section{Comparison between image reconstruction techniques} \label{DaCC} 

The reason for doing a comparison of post--facto reconstruction techniques is to understand if the observed rapid events are an artefact of a particular technique or are more evident using certain methods (specially addressing the apparent ultra--fast events). More technical comparison of both techniques is discussed in \citet{2011A&A...533A..21P} where they have compared properties of images from the speckle reconstruction and MOMFBD reconstruction using images collected from the G\"{o}ttingen Fabry--P\'{e}rot Interferometer (FPI) in Fe~{\sc i} (6302.5\AA) for spectro--polarimetry and in the blue continuum (4313\AA) and Ca {\sc ii} H (3968.5\AA) for imaging. For the spectro--polarimetric data they concluded that speckle deconvolution and MOMFBD were comparable but for imaging data they found MOMFBD reconstructed images were slightly better (with higher contrast) and more stable. In their spicule work (specially addressing the ultra-fast events), \citet{2012ApJ...755L..11J} found that both methods gave similar results whereas \citet{2014ApJ...785..109L} showed that the speckled data performed better than the MOMFBD data. They considered images reconstructed at different cadences of 1s and 5s for the MOMFBD and speckled techniques respectively. Further they found that the fibrils observed using both methods were very similar. Here we discuss in detail how we compared the reconstruction techniques.

\subsection{Observational results} 

In Fig. \ref{comp}, we show three consecutive normalized images taken in the blue wing of H$\alpha$ at --774 m\AA\, for speckle, ML and MH datasets as described in Sect.~\ref{dR}. The cadence for speckle and ML are both 3.9s, while for MH it is 0.7s. The  location of the region-of-interest is shown by the box outlined in red in Fig.\ref{sROI}. Here, the speckled and  ML images presented in the top two rows are reconstructed for the same time whereas for MH, we have used the nearest time i.e a few milliseconds apart. The yellow arrows indicate the location of a feature of interest. We observe this spicule in all three cases and the physical parameters such as length and width appear the same. The intensity change for this feature is roughly 15$\%$ (see Fig.~\ref{case1_wl} later). Another feature is tracked in Fig.~\ref{compmissing}, but we see the complete evolution only in the speckled data. The initial faint stages are missed in the MOMFBD data. When it comes to evaluating physical parameters such as length and width, we found that all the methods showed similar results even though ML and MH have different number of frames. However, Fig.~\ref{compmissing} clearly shows differences in the detected structures using all techniques. Thus we can conclude that there is an observable difference between MOMFBD and speckle images in terms of definition of structures and brightness/depth levels but it is not significant enough to impact upon the measurements of typical parameters (such as height, width and lifetime). This was also found by \citet{2012ApJ...755L..11J} and \citet{2014ApJ...785..109L}. Each type of constructed image contains the same structures where the only real variance is the resolution of such structure. 

In the Fig.~\ref{Hist_comp_44}, we have compared intensity histograms for MOMFBD and Speckle for equal number of time samplings. MOMFBD is sensitive to contrast changes. We reconstructed MOMFBD for 44 frames using both CRISP--T and CRISP--R at --774 m\AA\,(same frames as Speckle), plus wideband (WB) as input objects.


\section{Data analysis and statistics}\label{DnS}
\begin{figure*}
\centering
\includegraphics[scale=0.35]{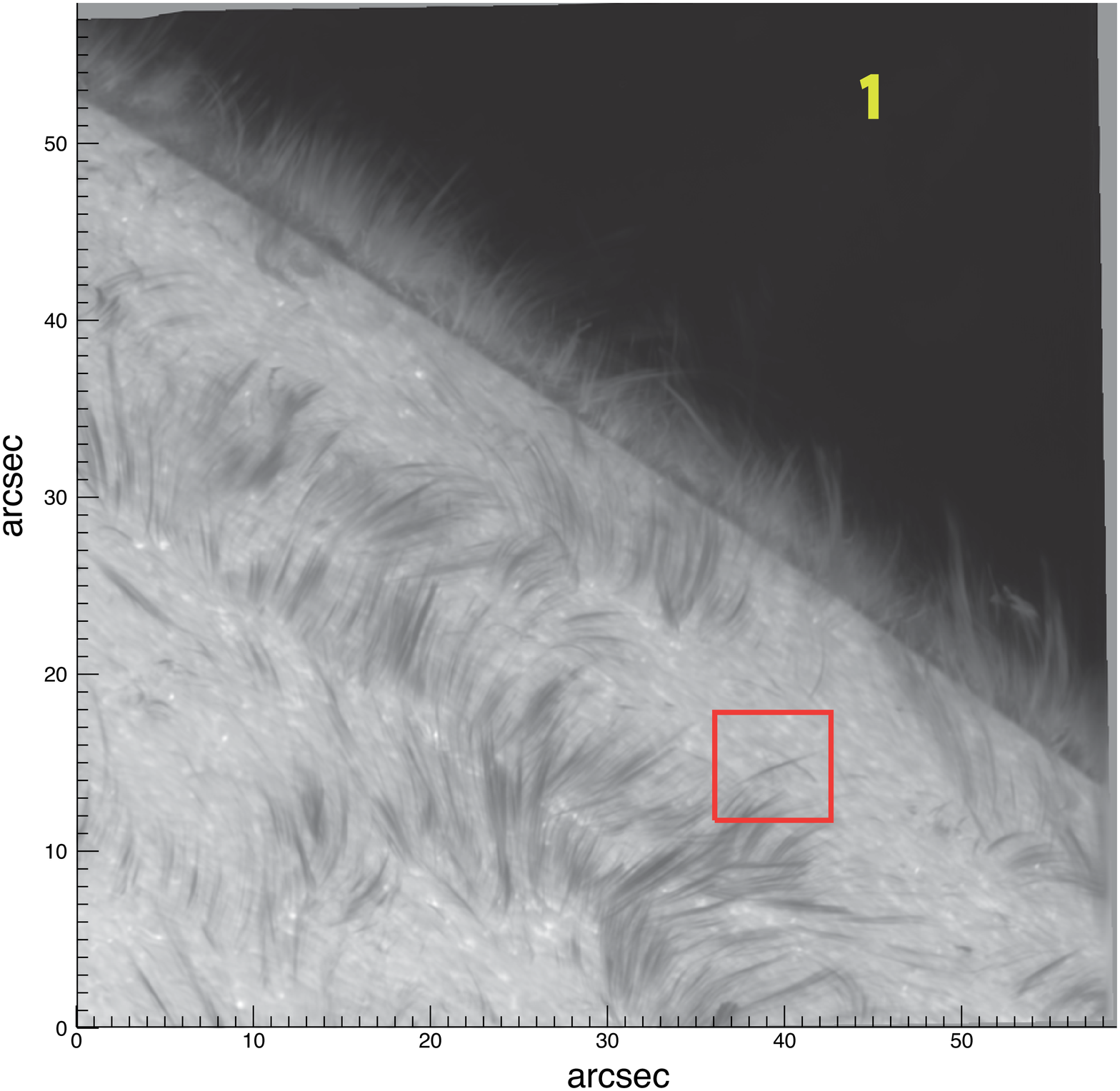}
\includegraphics[scale=0.7]{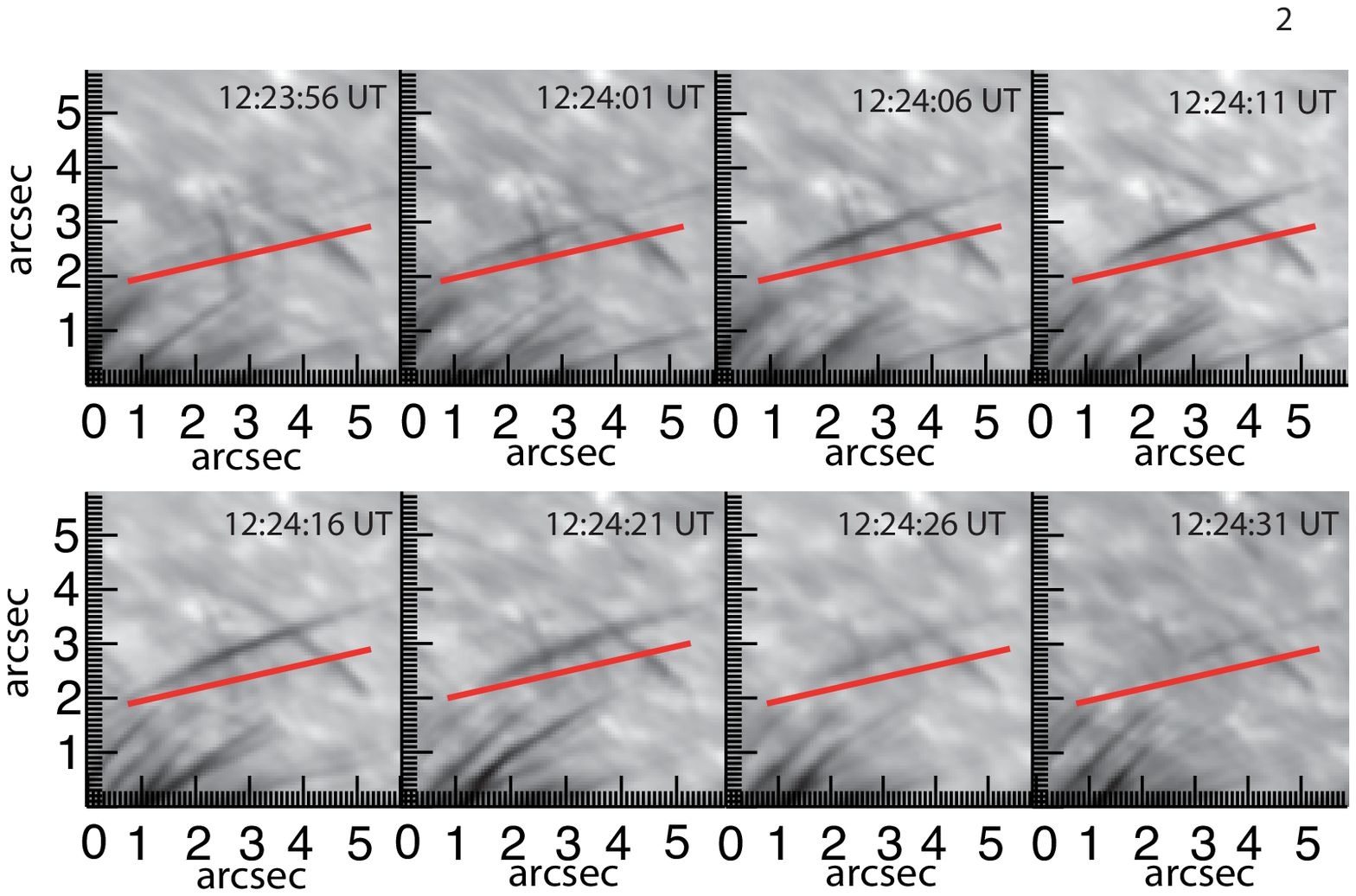}
\caption{Panel 1 shows the location of a feature observed on the 05 June 2014, which is exclusively reconstructed using the MOMFBD reconstruction technique. Panel 2 records the evolution of the feature with sub-panels corresponding to different sequential times starting at 12:23:56 UT then every 5 s until 12:24:31 UT. The solid red line is a reference line which allows the observer to see the lateral movement of this spicule.}\label{eg2}
\end{figure*}

\subsection{Intensity variations between different techniques}

The histograms of the normalized intensities for the first timestamp (08:07:22 UT) or first column of Fig.~\ref{comp} are presented in Fig.~\ref{Hist}. For this dataset, the standard deviation for speckle, MH and ML is 0.069, 0.053 and 0.0589 respectively. The broader intensity distribution for speckle indicates that this technique can be more helpful in identifying fainter structures. Furthermore, due to the broader (in Fig.~\ref{Hist}) intensity range of the speckle reconstructed images, it can be suggested that the method may pick up fainter structures better, a trend seen in Fig.~\ref{Hist_comp}. Here the histogram in Fig.~\ref{Hist_comp} is constructed using intensities corresponding to the images represented in the first column of Fig.~\ref{compmissing} corresponding to the time 07:43:53 UT, where the feature is identified only in the speckle technique. 
On comparing histograms for both cases (Fig.~\ref{comp} and Fig.~\ref{Hist_comp}), we can conclude that the missing structures in Fig.~\ref{compmissing} are because this feature could not be resolved in the MOMFBD data due to the intensity contrast. The rise phase of the event was missed due to the sequencing of the scan or the method of reconstruction used at the two wavelength positions. Since we use 44 images at two wavelength
positions for speckle and only 8 MOMFBD images, if a structure is present in one
image while performing speckle, the signature gets repeated for 44 images as speckle
depends on selecting a single PSF, where as it could be missing during MOMFBD
reconstruction. This also agrees with the low intensity ranges found in MOMFBD method. The reason being we simply have 5.5 times more frames for speckled images than MOMFBD at the compared wavelength position. While considering the full FOV of the image we also found that speckled data had a better intensity coverage compared to the MOMFBD data. We have then compared the contrast for the image reconstructed with Speckle and MOMFBD using the same number of time-slices. In the intensity histogram of Fig.~\ref{Hist_comp_44} we see that speckle (red histogram) still has a broader intensity range than MOMFBD (blue histogram), here the standard deviation is 0.066 for speckle and 0.053 for MOMFBD reconstructed image. We note that the difference between standard deviations is proportionally much smaller. If MOMFBD's mean intensity were the same as speckle's, for the same standard deviation, the MOMFBD's contrast would be 0.799065. However, for the analysis of various motions and properties of the structures (such as length or width), the differences in contrast/photometry do not seem to be important.   
\begin{figure*}[t]
\centering
\hspace*{-0.5cm}
\includegraphics[scale=0.55]{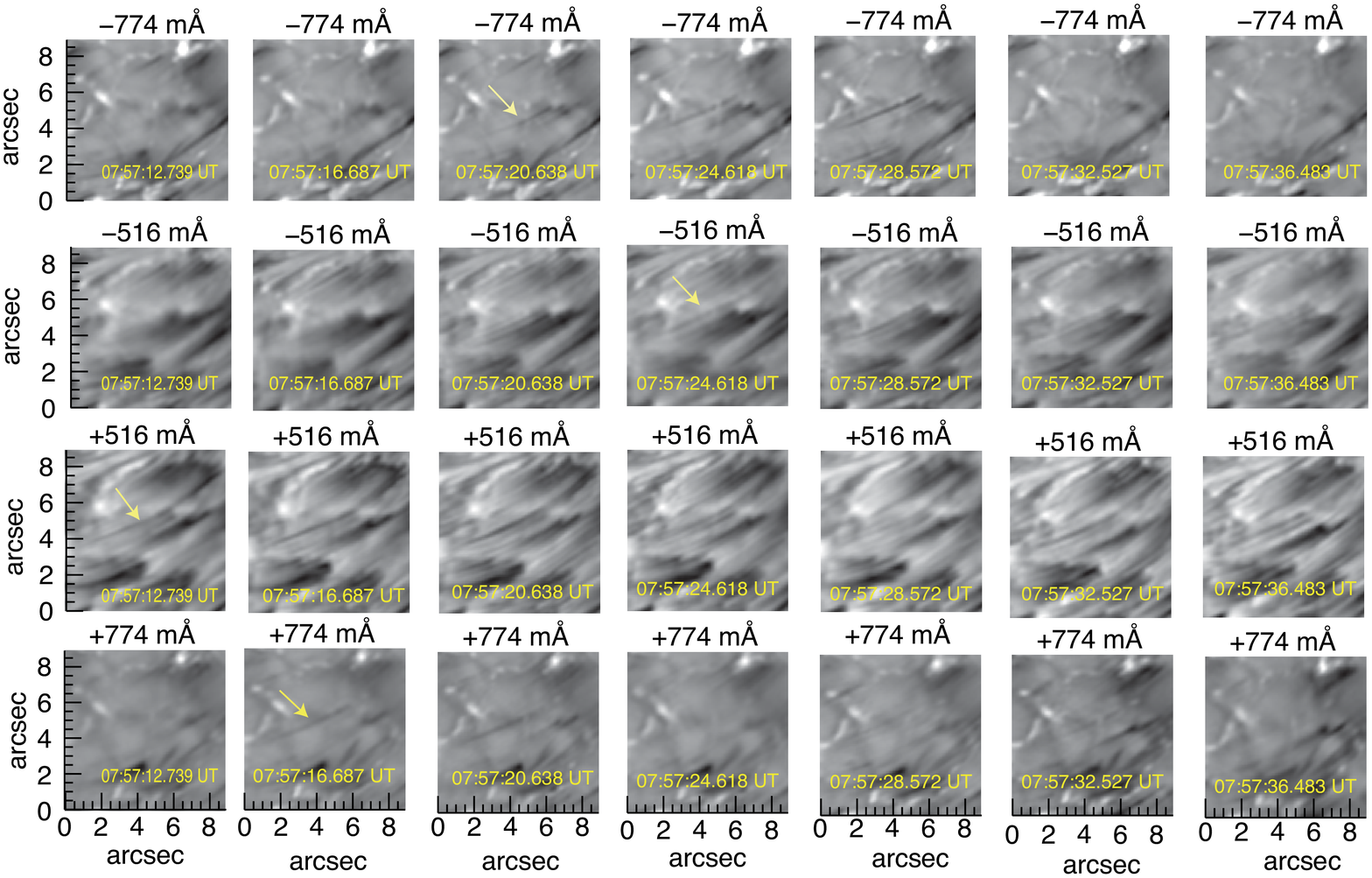}
\includegraphics[width=0.6\textwidth , height=0.25 \textheight]{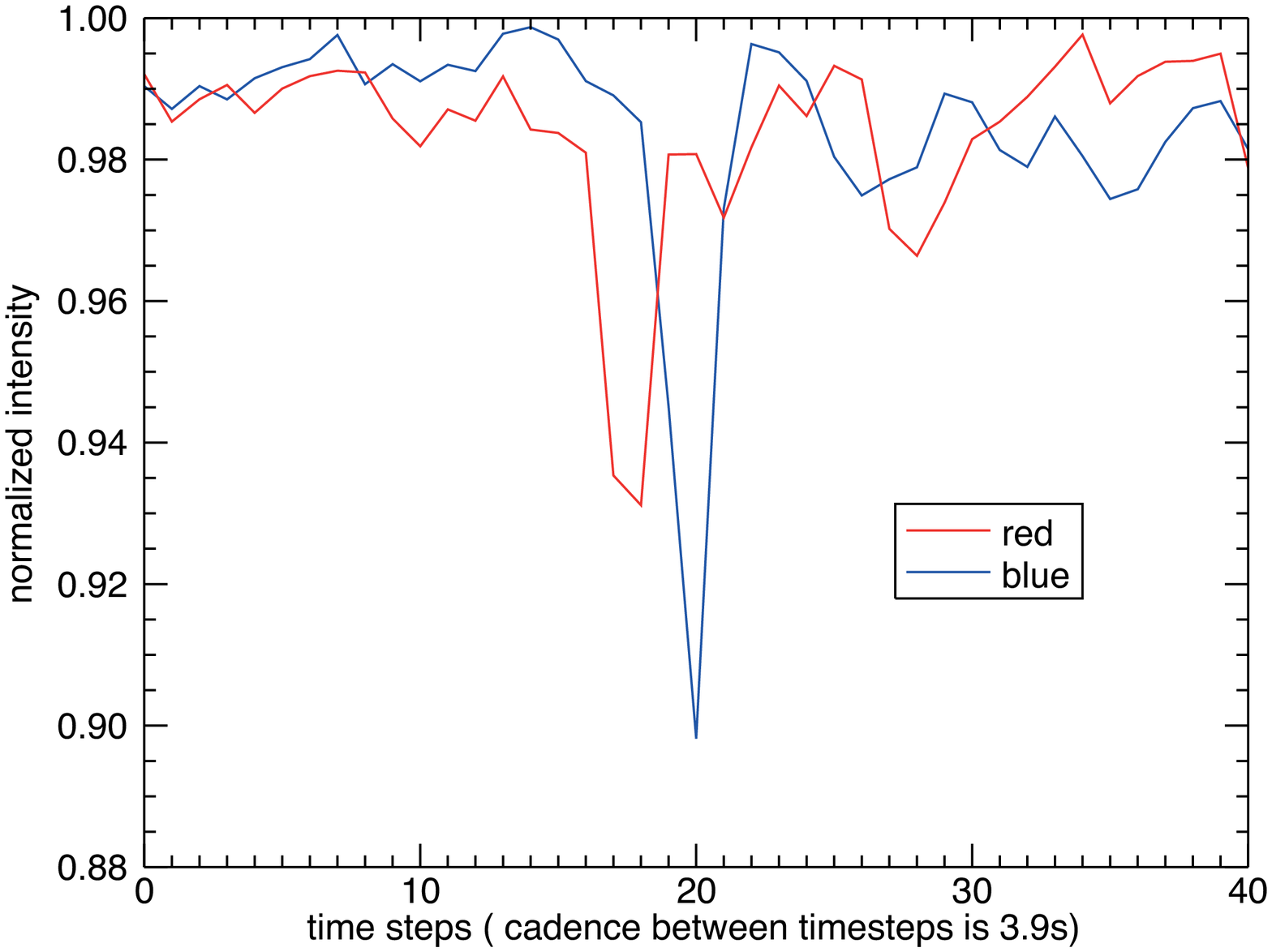}
\caption{Evolution of a spicule observed in four different wavelengths $\pm$ 774 m\AA\ and $\pm$ 516 m\AA\ in the 10 June 2014 dataset. The columns are divided into 7 panels starting at the time 07:57:12.739 UT with steps of 3.9s until 07:57:36.483 UT. The yellow arrow shows the location of the spicule first observed in that wavelength. The image intensities are normalised. The lower panel shows a normalized light curve of the feature indicated by the yellow arrow in the top panel. These light curves are plotted by summing intensity variations along ten positions along the length of the spicule.}
\label{case23_wl}
\end{figure*}  

\begin{figure*}
\centering
\vspace*{0.5cm}
\includegraphics[scale=0.45]{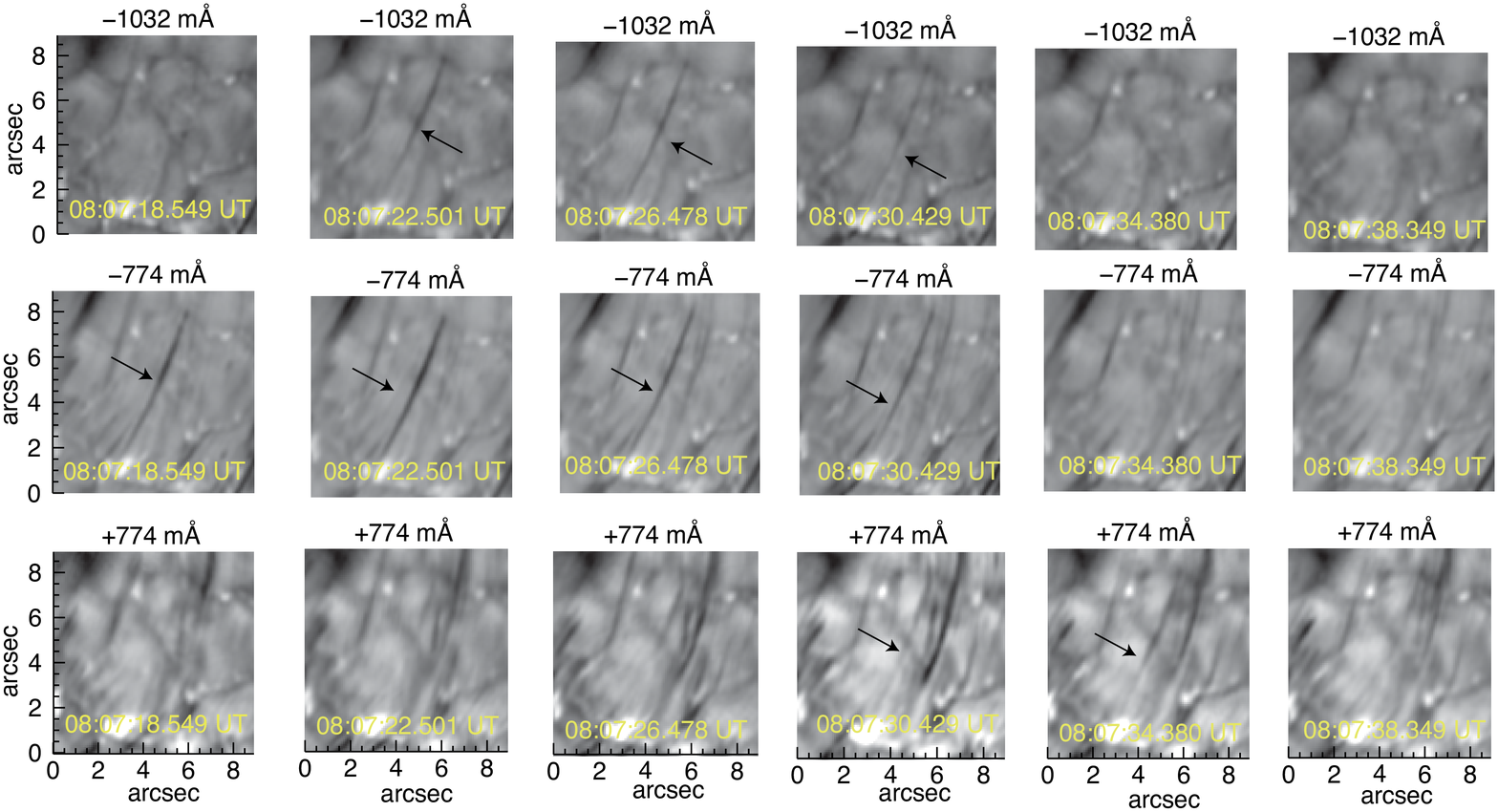}
\includegraphics[width=0.6\textwidth , height=0.3 \textheight]{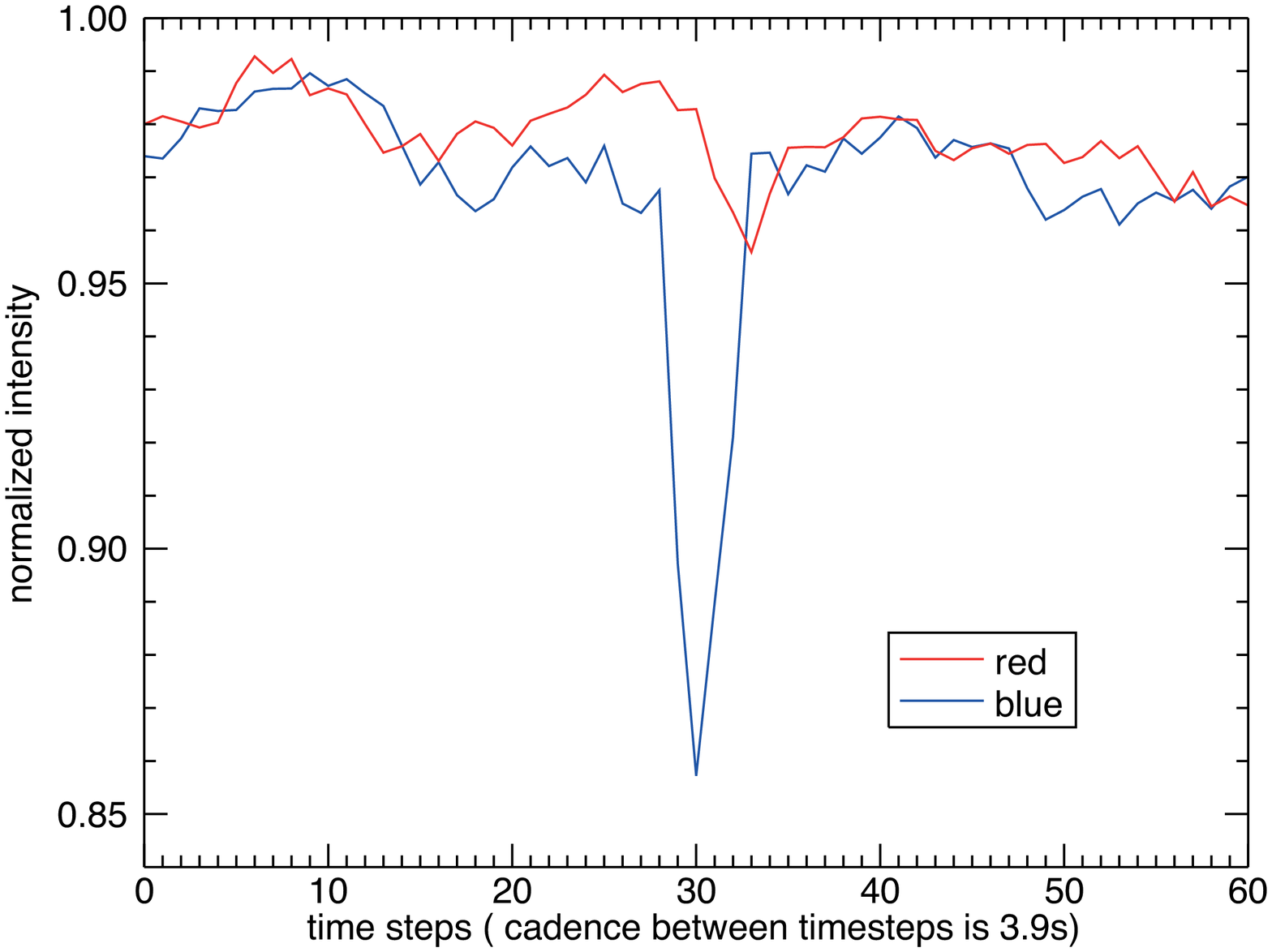}
\caption{ Example of a spicule from the 10 June dataset displaying its signature at different wavelengths, here at -1032 m\AA\, and $\pm$ 774 m\AA. The evolution is divided over six columns with times corresponding starting at 08:07:18.549 UT to 08:07:38.349 UT, with 3.9s difference in each columns. The black arrows represent the times when the spicules were observed at that wavelength. The image intensities are normalized. The lower panel shows a normalized light curve of the feature indicated by the black arrow. These light curves are plotted by summing the intensity variations along ten positions along the length of the spicule.}
\label{case1_wl}
\end{figure*}

\textbf{\subsection{Statistical analysis of spicules}}
Features or spicules analysed here are seen as dark absorption structures in the H$\alpha$ wings at $\pm$516 m\AA, $\pm$774 m\AA\, and $\pm$1032 m\AA\,. For the 10 June dataset, we used both the speckle images and MOMFBD images for identifying these features in the blue wing of the H$\alpha$ line profile at --774 m\AA\ while for the 5 June dataset we used only MOMFBD reconstructed images. For the apparent ultra-fast spicules we searched for events that emerge in one time-step (3.9s for 10 June and 5.0s for 5 June datasets). These are relatively faint features as the change in intensity of the features against the background intensity lies between 5$\%$ to 40$\%$ (see the light curves plotted in Fig. ~\ref{case23_wl} and Fig.~\ref{case1_wl}). The light curves plotted in Fig.~\ref{case23_wl} and Fig.~\ref{case1_wl} are constructed by taking the intensity at 10 positions along the length of the feature and then averaging the value over time, thus giving intensity evolution with time.  

For the 10 June dataset, we had 38 mins of data which resulted in the detection of 23 of these apparent ultra-fast features. For the 5 June dataset, we had 50 mins of data which resulted in the detection of 17 features. We found that 25$\%$ of these features re-appeared at the same location several minutes later. In Table.~\ref{A1} and Table.~\ref{A2}, we give full details for all the features detected. In our observations most features were detected to have a blue wing signature (78$\%$) whereas (60$\%$) had a signature in the red wing. Several features showed both a red and blue shift, although in most events there is a time delay in their appearance whether in the blue or red. From the total number of features observed 75$\%$, show signatures in more than one line position viz. $\pm$1032 m\AA, $\pm$774 m\AA\ and $\pm$516 m\AA, which correspond to approximate Doppler velocities of $\pm$ 47 km s$^{-1}$, 35 km s$^{-1}$ and 24 km s$^{-1}$.

Some examples of these features are represented in the top panel of Fig.~\ref{case23_wl} and Fig.~\ref{case1_wl} for the 10 June dataset and panel $''$2$''$ of Fig.~\ref{eg2} for the 5 June dataset. For example, in Fig.~\ref{case23_wl} we first note the appearance of a feature at the +516 m\AA\ position in the red wing, it then appears at +774 m\AA\ and after a further 3.9s the feature shows a signature in the blue wing. The yellow arrow in Fig.~\ref{case23_wl} indicates the location of the feature when it first emergences in the CRISP FOV. In another example shown in Fig.~\ref{case1_wl}, a feature is first detected in the blue wing of H$\alpha$ at --774 m\AA, after a further 7.4s it is seen in the red wing position +774 m\AA, here the black arrow indicates the first appearance of the feature. The light curves (see bottom panel of Fig.~\ref{case23_wl} and Fig.~\ref{case1_wl} corresponding to the features in the above examples) confirm the temporal offset between the appearance of the features. This behaviour indicates that these features do not have any preferred direction of motion. We also note that in most cases they show signatures at multiple line positions and that their lifetime varies between the different line positions. Often we find that features detected in the far-red and far--blue wings have a shorter life-span, with those detected at $\pm$ 774 m\AA\ lasting the longest. Again see the example in Fig.~\ref{case1_wl}, here the top row shows the feature's evolution at --1032 m\AA\, seen for 11.1s whereas it is seen for 14.8s at --774 m\AA\,. This gives an indication of an accelerated Doppler shift.  

\begin{figure}[h!]
\centering
\includegraphics[scale=0.26]{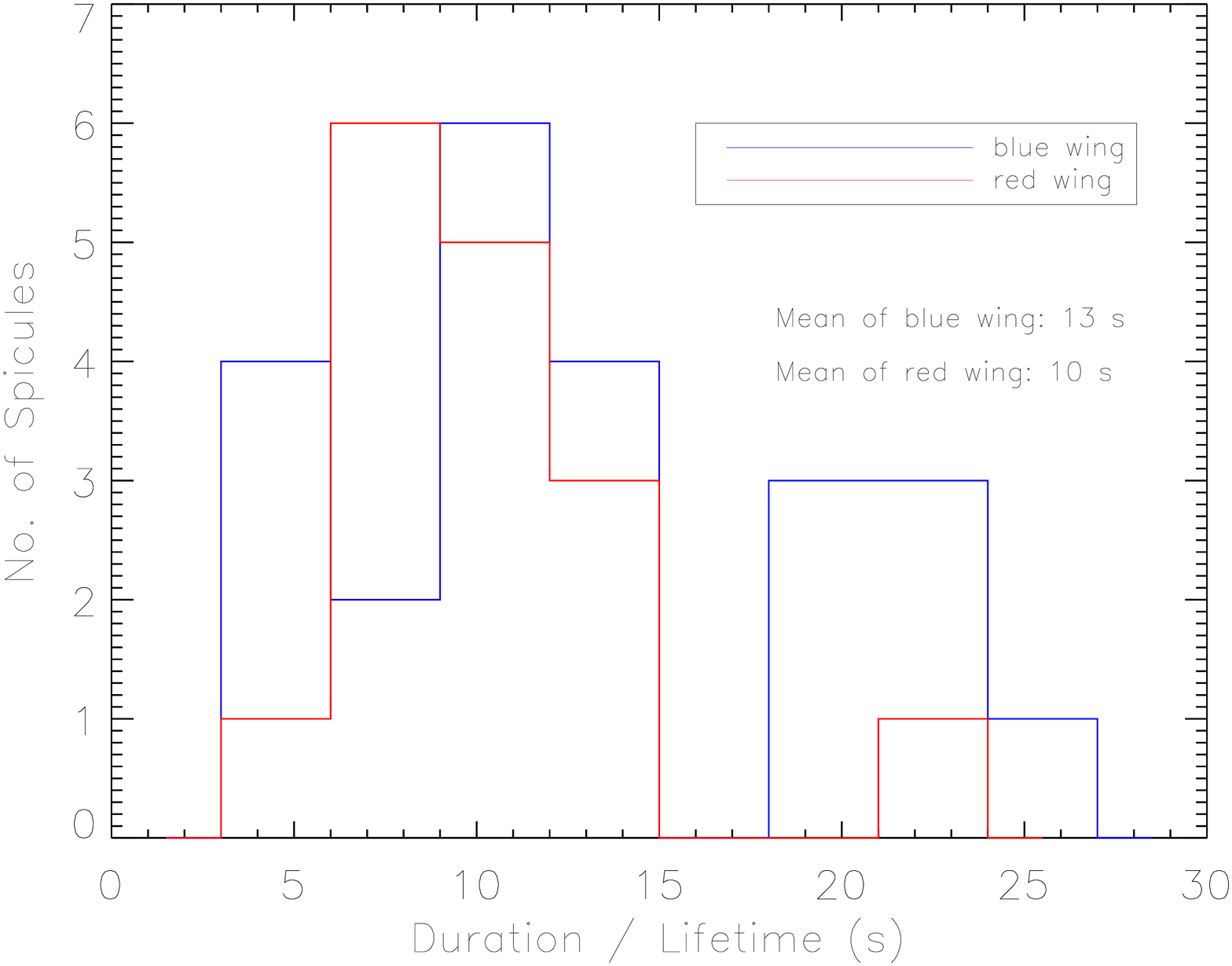}
\includegraphics[scale=0.26]{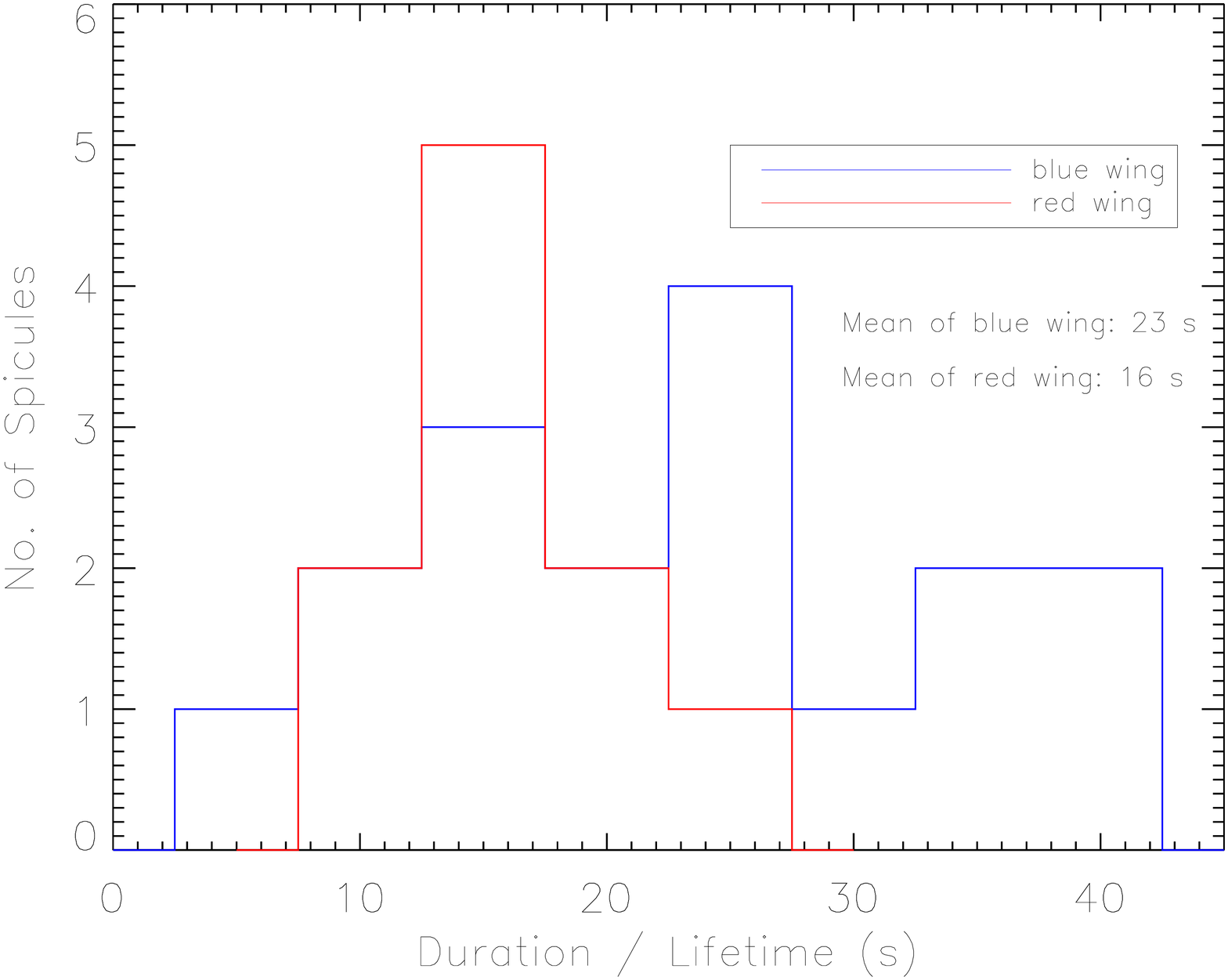}
\caption{The number of detected apparent ultra-fast spicules v. lifetime for the 10 June 2014 (top panel) and 5 June 2014 (bottom panel).} \label{lifetime}
\end{figure}

\begin{figure}[h!]
\centering
\includegraphics[scale=0.26]{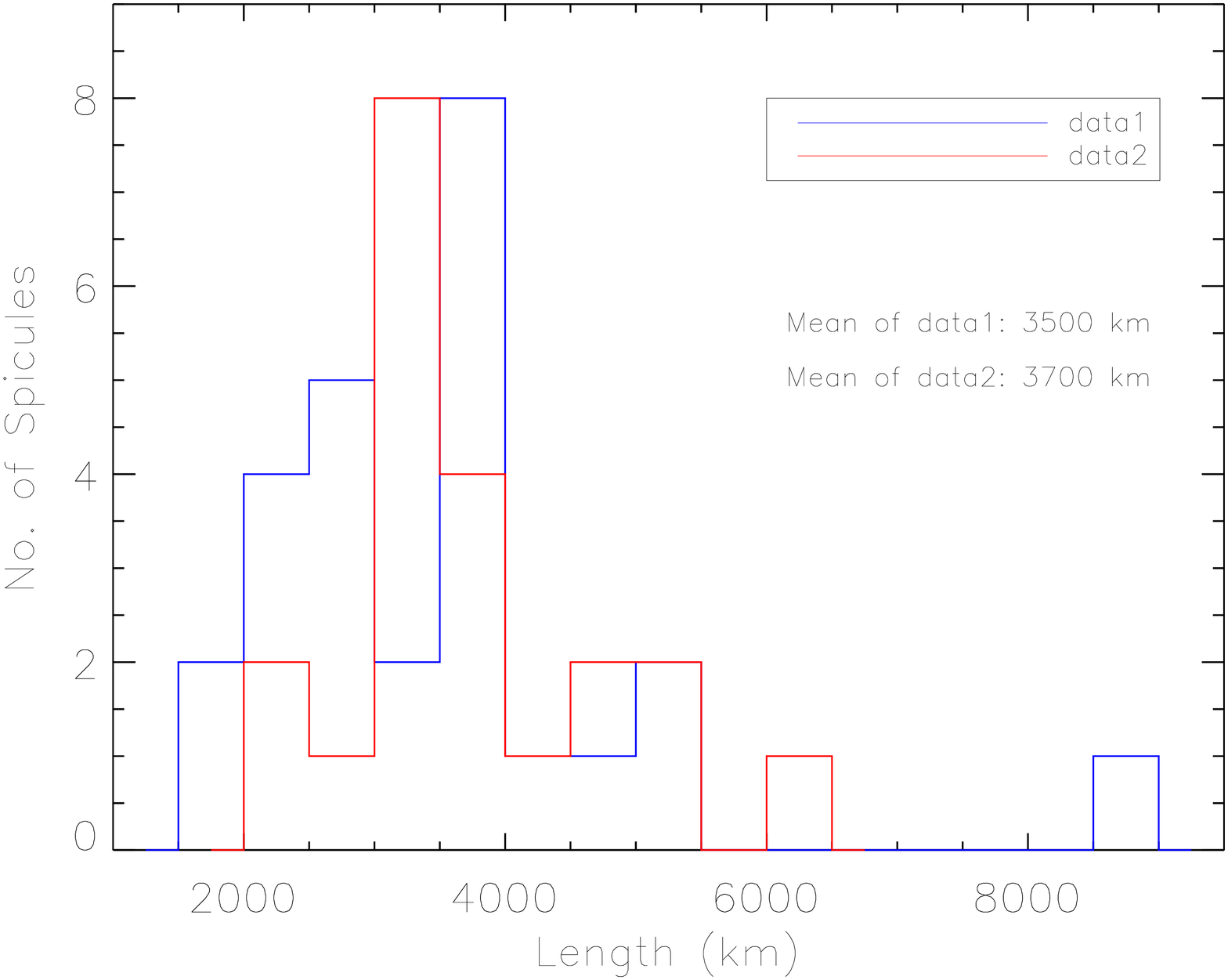}
\includegraphics[scale=0.26]{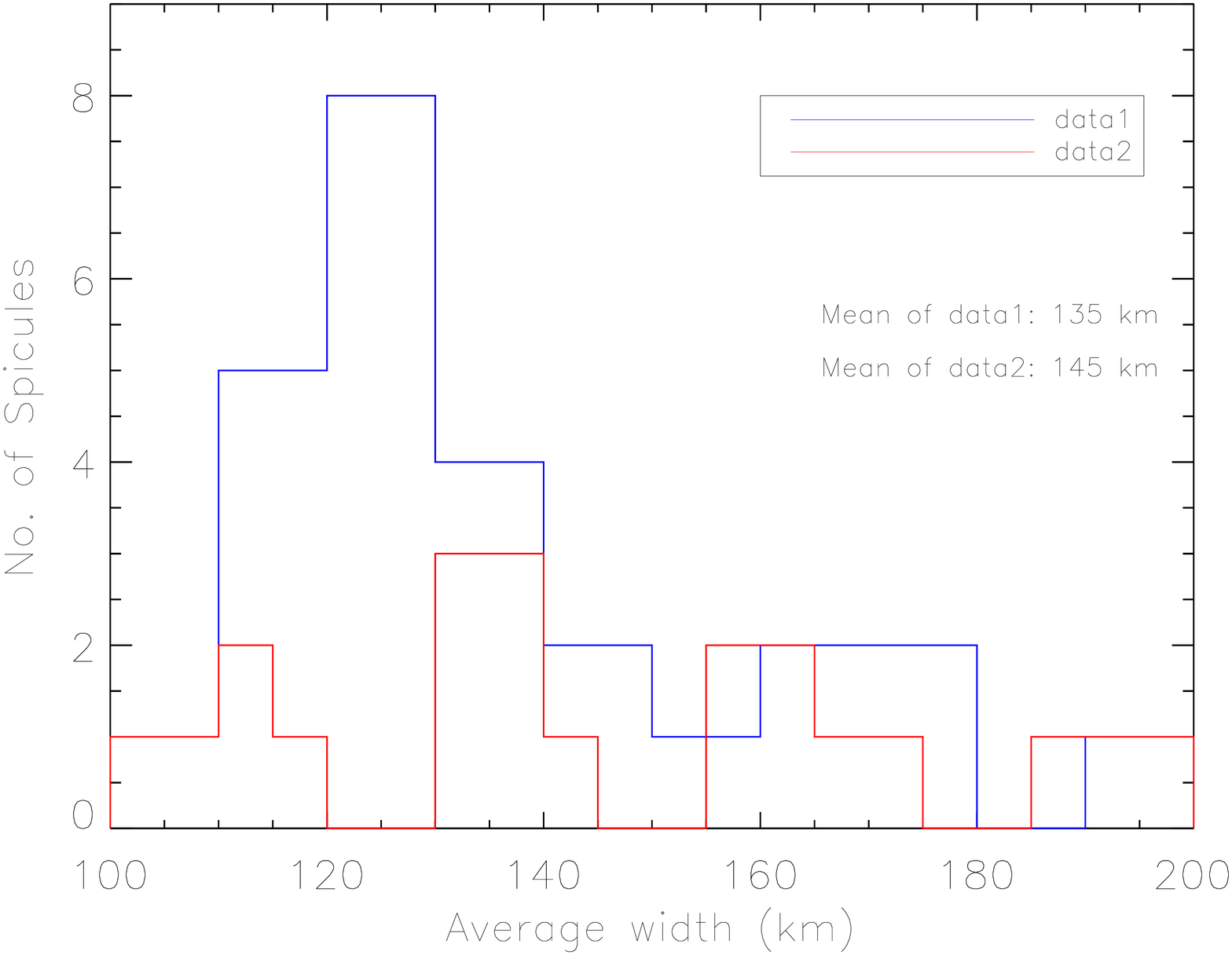}
\caption{Statistics of physical parameters. The bottom panel represents the number of features v. the average width and the upper panel represents the number of features v. maximum length / height of the spicules. The legend data1 represents the 10 June dataset and legend data2 represents the 5 June dataset.}\label{stats}
\end{figure}

In order to quantify some physical parameters, we define the lifetime as the time between the sudden appearance of the feature and its complete disappearance (i.e. the background level is reached) at that particular line position. The average lifetime for the 10 June dataset is 11.9s, while for the 5 June dataset it is 19.5s (see Fig.~\ref{lifetime}). This difference could be a function of both the cadence and the viewing angle. The feature with the longest duration only lasted 25.9s in the 10 June dataset and 40s in the 5 June dataset. In both datasets, the features in the red wing had a shorter lifespan than those in the blue wing. The length or the maximum height for the features is the observed maximum length during the evolution of the feature. The longest feature present in the 10 June dataset was 8600 km (with $\pm$ 82 km error) while for the 5 June dataset it was 6400 km (with $\pm$ 82 km error, also note that we have not taken into account the uncertainty in the projection of the spicules, which could be responsible for the smaller length). For the width, we measured the width at three positions along
the feature using CRISPEX, finding 191 km and 196 km (with $\pm$ 41 km error) for the 10 June and 5 June datasets respectively (see Fig.~\ref{stats}). These histograms represent the number of detected apparent ultra--fast spicules v. length and average width. On comparing the width to length ratio we see that these features are very long and narrow, very similar to spicules \citep{2012ApJ...759...18P} and RBEs \citep{2015ApJ...802...26K}. 

\begin{figure}
\centering
\hspace*{-1.1cm}
\includegraphics[scale=0.26]{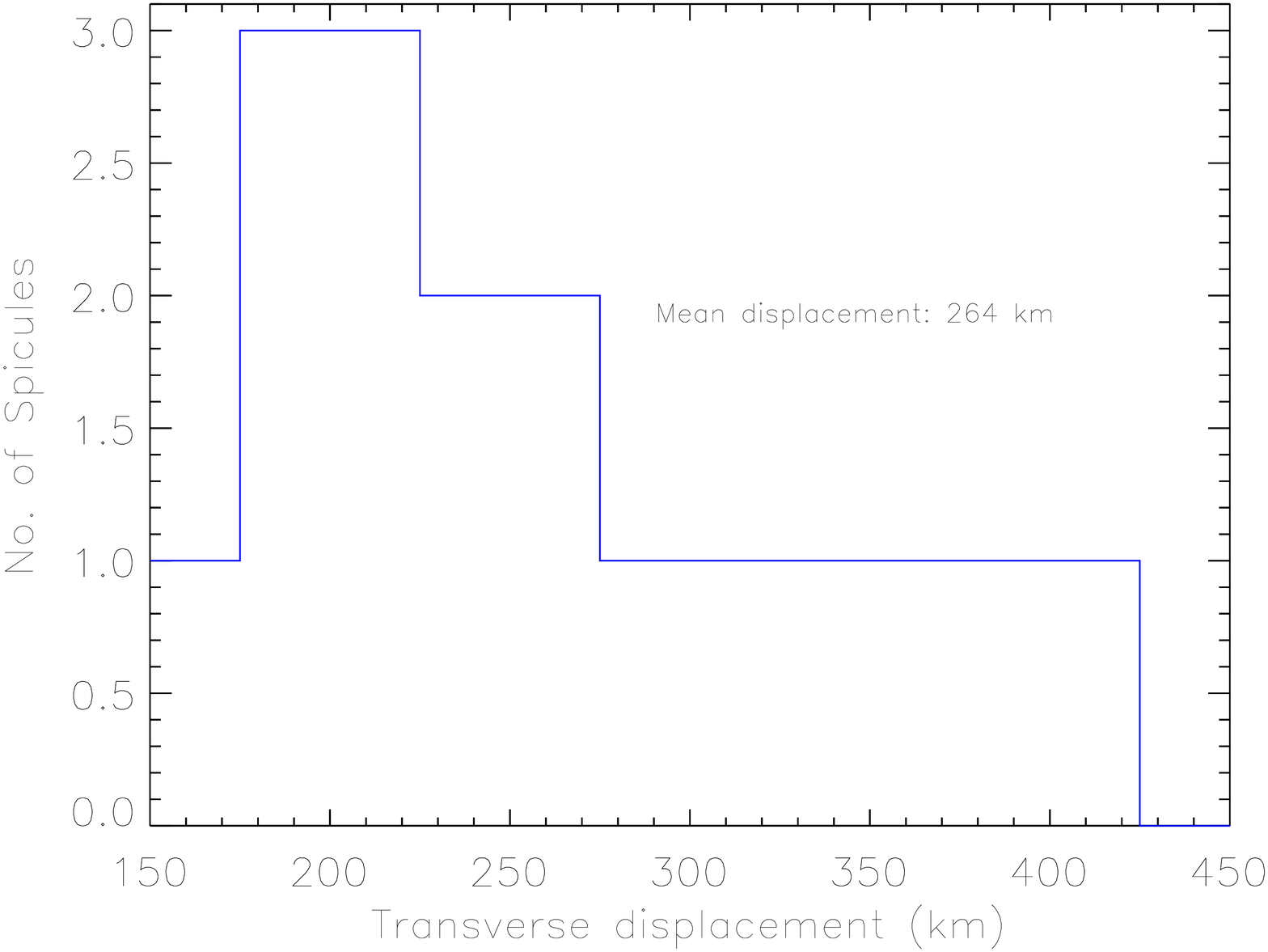}
\hspace*{-1.1cm}
\includegraphics[scale=0.26]{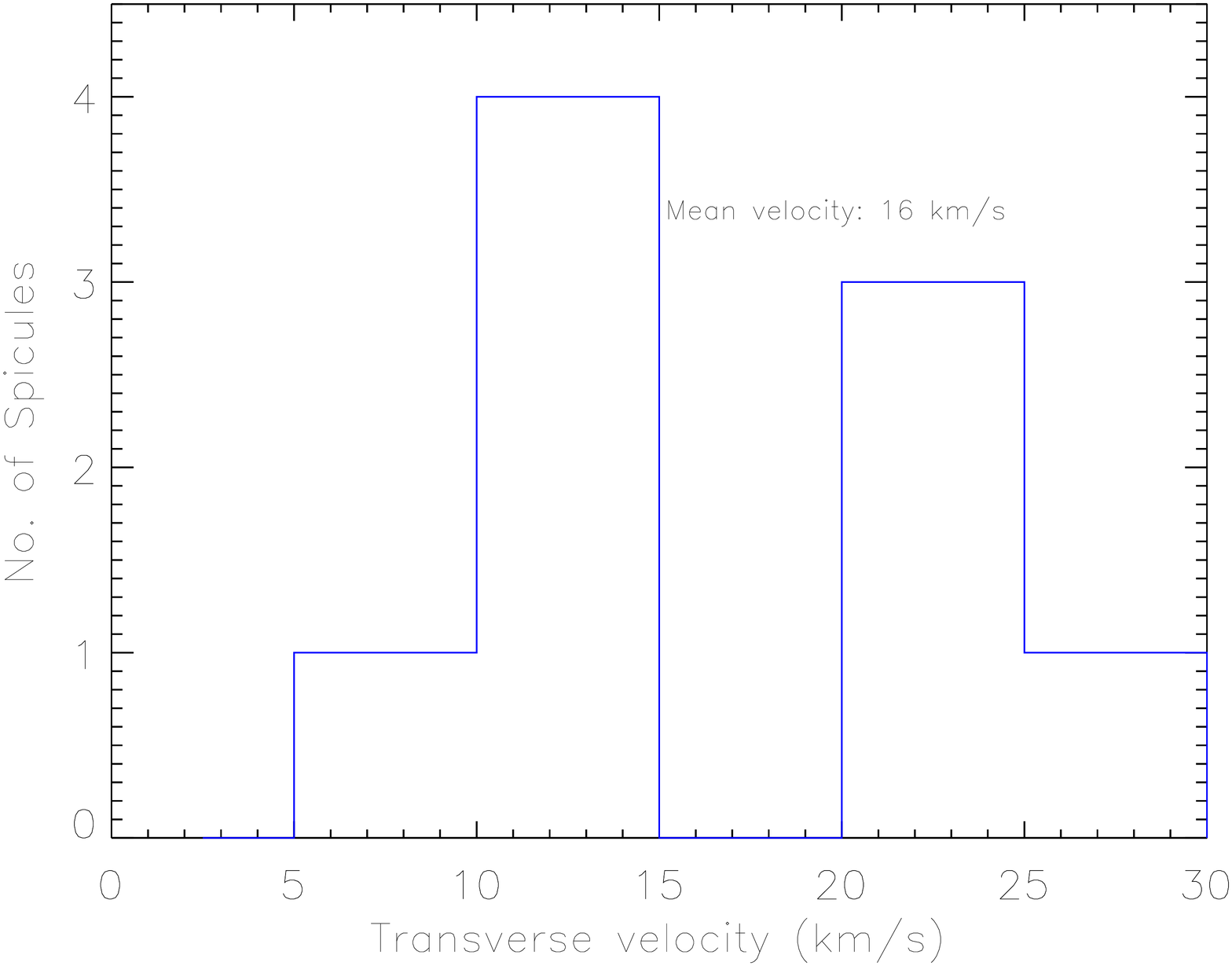}
\caption{Transverse displacement (top panel) and velocity plots (bottom panel) for the features}
\label{transverse}
\end{figure}

\begin{figure}[h!]
\centering
\includegraphics[scale=0.375]{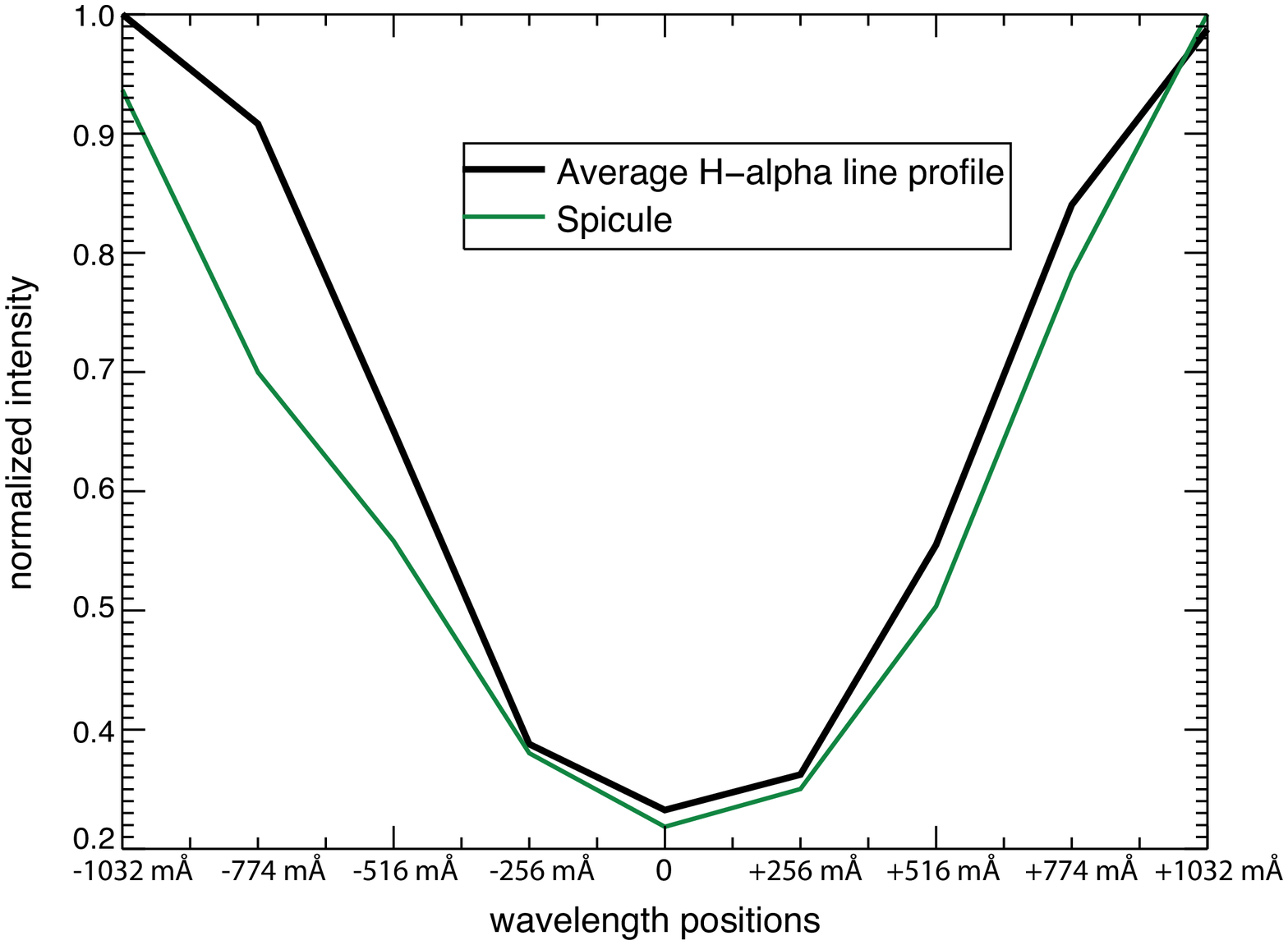}
\includegraphics[scale=0.375]{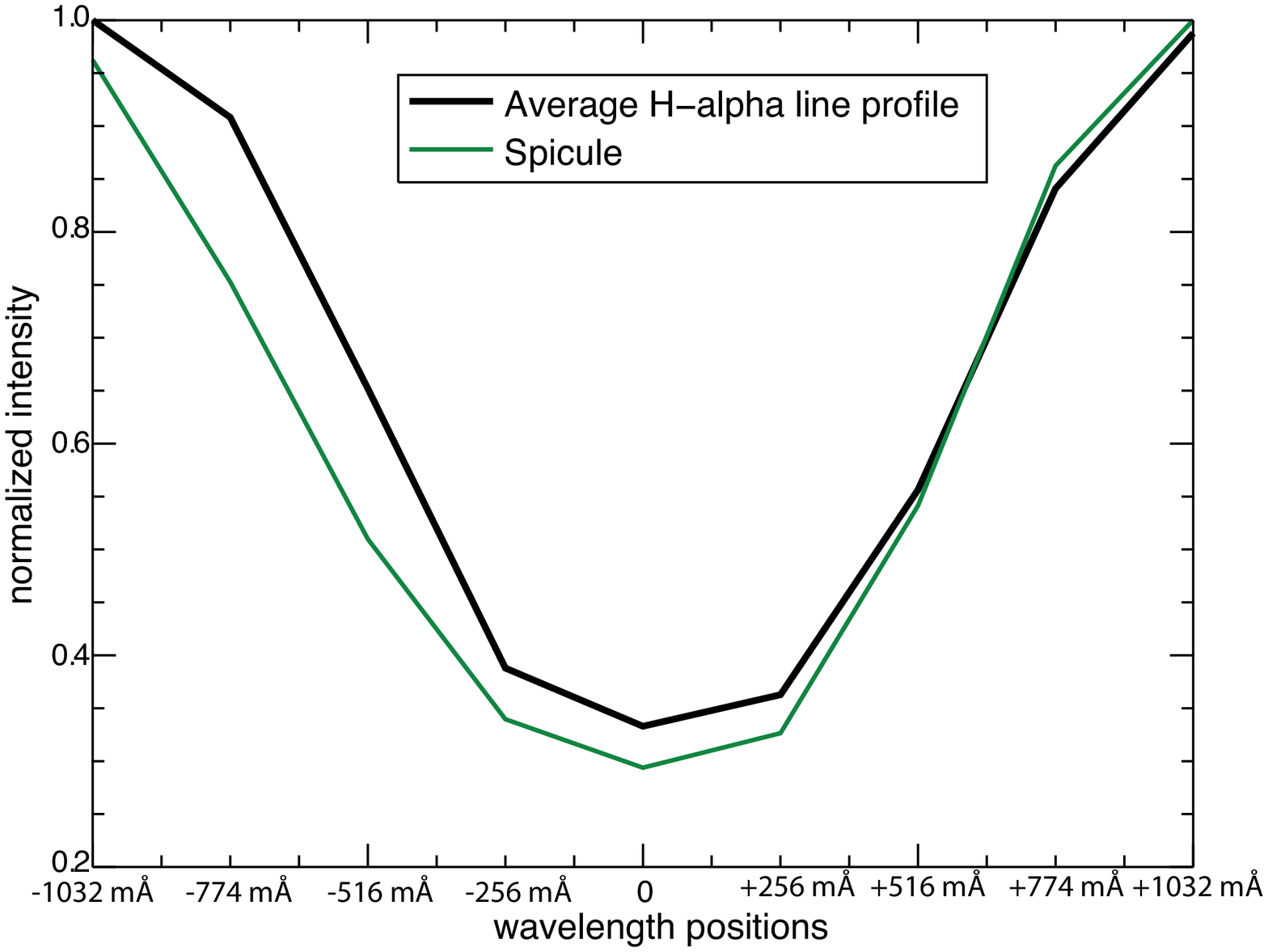}
\includegraphics[scale=0.375]{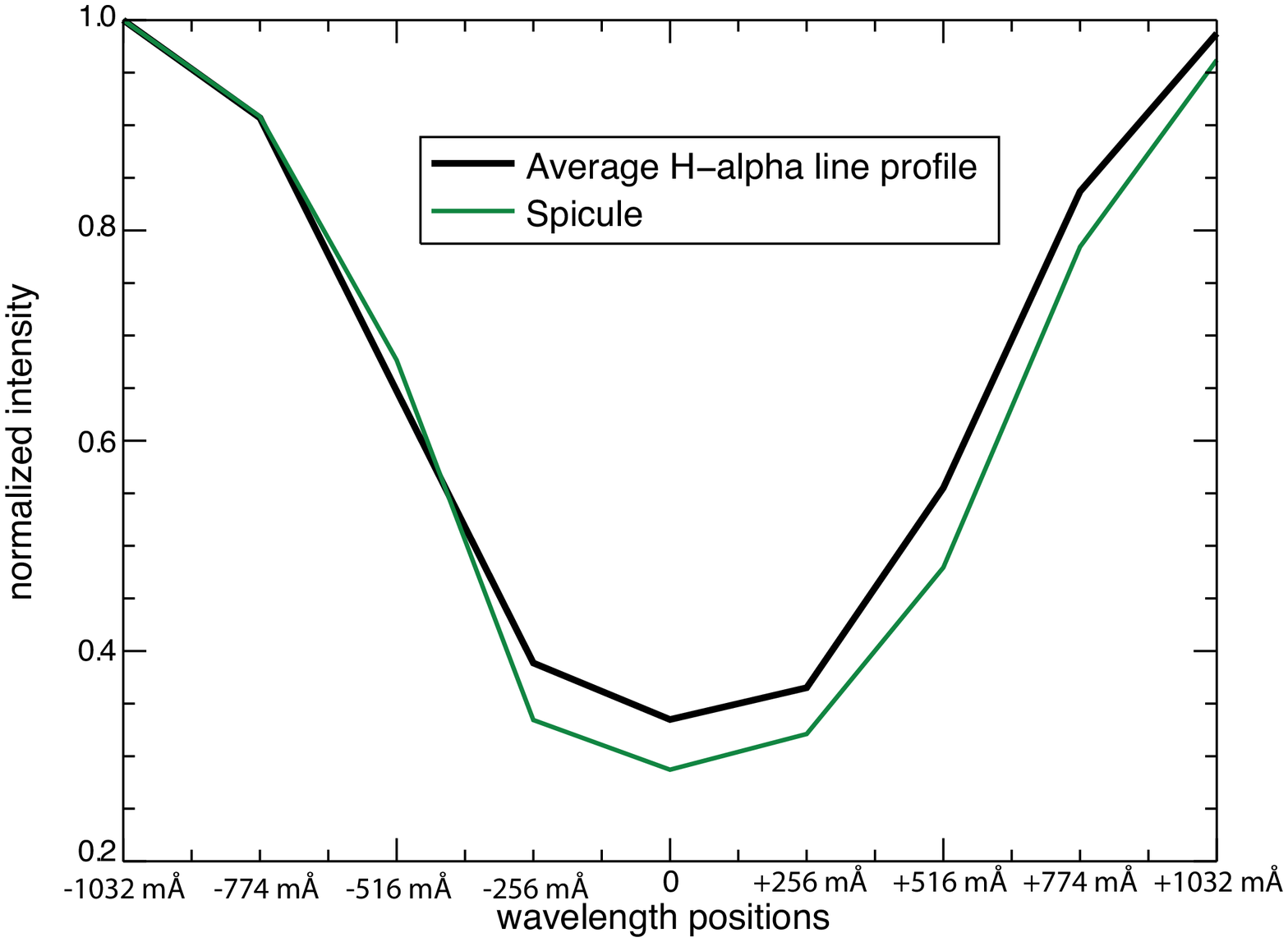}
\caption{H$\alpha$ line profiles of three spicular--type events. Top panel: where the feature showed a relatively strong blue--wing signature, this feature shows is similar to that shown in Fig.~\ref{case23_wl}, a canonical example of RBE that re--occurs in different line positions. Middle panel: A feature which shows a blue--wing signature with a broadened line--core. Bottom panel is a feature with strong signature in red--wing positions again with a broadened line--core. This can be considered as a RRE.}\label{lineprofiles}
\end{figure}

\begin{figure*}[htb]
\centering
\includegraphics[width=0.83\textwidth , height=0.53 \textheight]{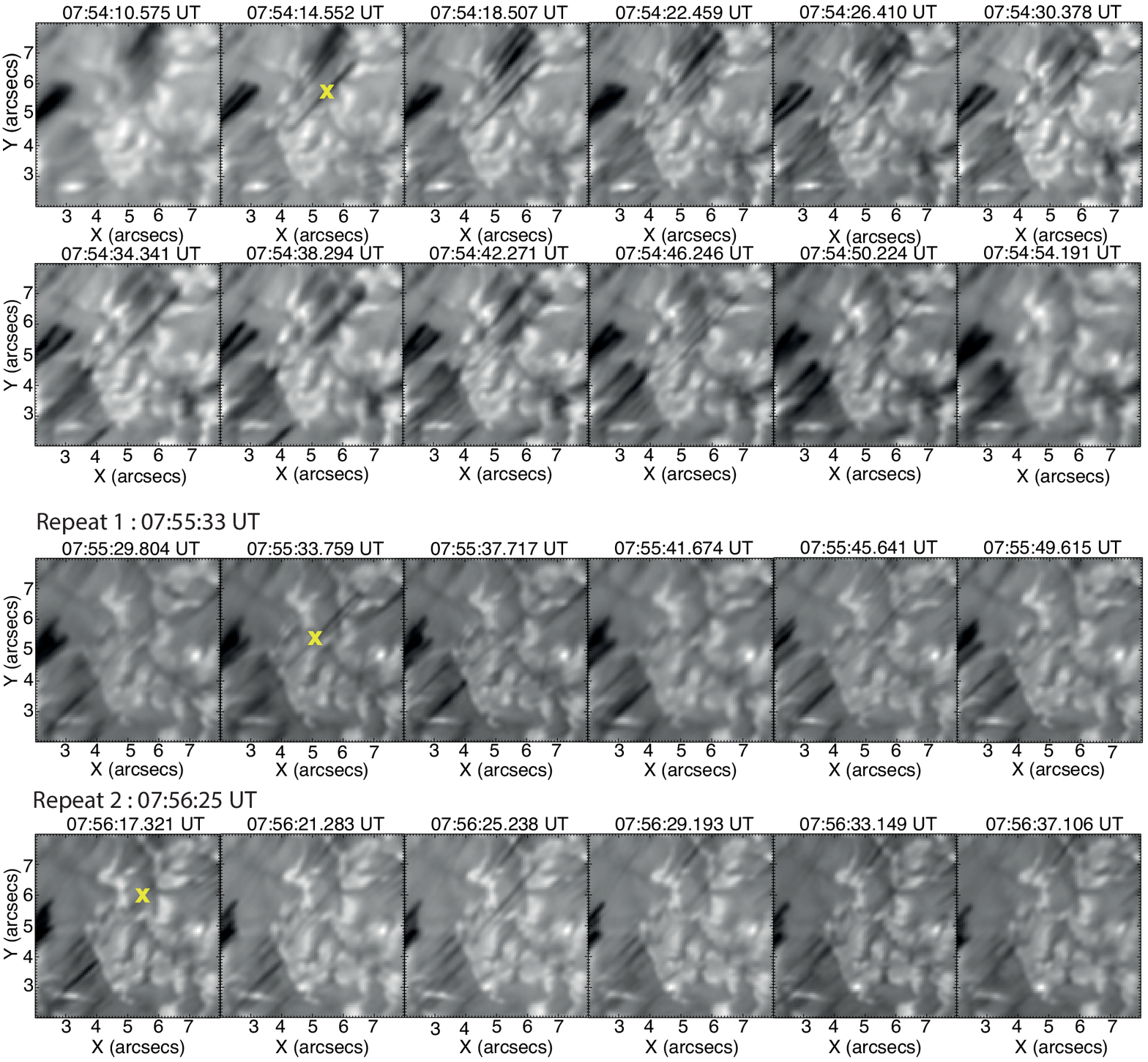}

\caption{An example from 10 June 2014 where a feature first appears at 07:54:10.575 UT then seems to disappear out of our FOV at 07:54:54.191 UT. It re-emerged again at 07:55:29.804 UT and 07:56.17.321 UT, as indicated by the yellow cross.}
\label{repeat}
\end{figure*}

Furthermore $\approx$25$\%$ of the features in our datasets show lateral/transverse shifts. In most cases we observed shifts in only one direction. An example of a lateral shift during the evolution of a feature observed on 05 June 2014, is seen in panel 2 of Fig.~\ref{eg2}. The lifetime of this feature is 35s which is best seen from 12:24:01 UT to 12:24:26 UT, each subfigure being separated by 5.0s. The red-line in the panel is a reference line. This feature seems to show transverse movement away and towards the reference line. Note that the line-of-sight correction is not applied to the feature. Further statistics of transverse motion are represented in Fig.~\ref{transverse}. Here we represent statistics of features mainly observed in the 05 June 2014 dataset as there were only three cases of transverse motion observed in the 10 June dataset.The maximum transverse displacement is defined by tracing the shift using the CRISPEX software. We verify this by adding a slit across the feature at three positions and tracing the distance using the TANAT software. We then take the average of the distance. The transverse speed of the feature is calculated by considering the maximum displacement of the spicule from the starting position and time the feature takes to reach the maximum displacement. We found the mean displacement for transverse motion was 264 km and the velocity was 16 km s$^{-1}$. For the three cases that showed lateral motion for the 10 June dataset, the mean displacement is 134 km and the velocity is 36 km s$^{-1}$. The difference is most likely due to the viewing angle. We analyse these transverse motion in more details in an upcoming paper.

\vspace*{0.6cm}
\section{Discussion}\label{Diss}

\citet{2011ApJ...730L...4J,2012ApJ...755L..11J} and \citet{2014ApJ...785..109L} questioned the traditional interpretation that all spicules consists of a flux tube where plasma flows along the length. Instead \citet{2011ApJ...730L...4J,2012ApJ...755L..11J} suggested that in some instances spicules can form in sheet--like structures of chromospheric material. The features reported in this paper are similar to those reported by the above authors. If interpreted in terms of mass outflows, then velocities in excess of 1000 km s$^{-1}$ would be implied. We suggest that these observations can be interpreted as the movement of highly dynamic spicules in and out of the narrow 60m\AA\ filter used to observe in different parts of the H$\alpha$ line profile, i.e. LOS doppler velocity (which could be produced by the lateral motions and flows as well) is responsible of their appearance in the different wavelength positions across H$\alpha$ line profile. Also the simultaneous appearance of structures at different positions could be due to line broadening. In Fig.~\ref{lineprofiles}, we show the H$\alpha$ line--profiles corresponding to few features. The average H$\alpha$ profile were calculating the average profile for all pixels within the FOV. This profile was then compared against the H$\alpha$ line profiles for the features of interest. The top panel shows the feature that is canonical example of a RBE. The middle panel shows a feature with a strong blue--wing signature with a broadened line--core. This profile can be considered as a subset of RBEs. Finally, the bottom panel represents feature with strong signature in the red wing again with a broadened line--core. This particular profile can be considered as profiles of subsets of RREs.

In Fig.~\ref{eg2}, we observed lateral movement, where the spicule is seen to move laterally as compared to the solid red reference line. Fig.~\ref{case23_wl}, shows a spicule from the 10 June dataset, observed in different wavelengths $\pm$ 774 m\AA\ and $\pm$ 516 m\AA\,. The columns are divided into 7 panels corresponding to a start times of 07:57:12.739 UT with 3.9s time gaps until 07:57:36.483 UT. The yellow arrow shows the location of the spicule first observed in that wavelength. Another example is represented in Fig.~\ref{case1_wl} that shows a spicule from the 10 June dataset, observed in four different wavelengths $\pm$ 774 m\AA\, and far blue wing --1032 m\AA\,. The columns are divided into six panels starting at 08:07:18.549 UT with 3.9s time gaps until 08:07:38.349 UT respectively. Here the dark arrows shows the location of the spicule first observed in that wavelength. This example indicates the presence of highly dynamic motions, which we conclude as possible acceleration. This acceleration naturally arises from swaying motions of spicules \citep{2007Sci...318.1574D}. Furthermore, the absence of flow patterns in the time-slice diagrams of these features is indicative of the missing emerging and declining phases of these Doppler mass motions moving in and out of the 60 m\AA\,transmission filter. 

In addition to the dynamic appearance and disappearance of these features, we see repetitions or reemergence of these features in and out of CRISP FOV, confirming the presence of a dynamic feature at the same position. This could be a new feature or re-appearance of an old feature. Since they occur again in rapid succession without a visible flow pattern, these can be considered as the movement of flux tubes. Fig.~\ref{repeat} is an example of a feature which first re--emerges after 23s and after $\approx$ 60s from the first repeat. The top two rows of Fig.~\ref{repeat} represent the evolution of a structure which suddenly appears at 07:54:14.552 UT on 10 June 2014. The structure then seems to disappear out of our FOV at 07:54:54.191 UT. The next two rows show two repeats at the same position. All the panels are taken at --774 m\AA. The location of the feature is represented by a yellow $'$x$'$. We suggest that these spicule--type features are a subset of RBEs/RREs and that the rotating behaviour observed within them may be be associated with waves.

\subsection{Possible association with RBEs and RREs}
Looking at the physical parameters, such as length and width of these features, it is possible that the features observed here are a subset of the RBEs and RREs. \citet{2013ApJ...764..164S} presented a detailed statistical analysis using ultra--high cadence data (0.88 s) and found that some RBEs move sideways up to 55 km s$^{-1}$. Most of the features analysed by these authors showed rise velocities which varied between 40 and 127 km s$^{-1}$, however others suddenly appeared within a few seconds without an observable rise phase. A more recent statistical study by \citet{2015ApJ...802...26K} found that RREs and RBEs have similar occurrence rates, lifetimes, lengths, and widths. As found here, the features  appeared as upward (or downward) directed high speed jets/blobs, emerging  suddenly within a few time resolution elements ($\sim$1--5s). At the end of their lifetime, similar to type II spicules, the structures disappear extremely quickly. For Type-II spicules, this was  interpreted as being due to rapid heating to higher temperatures. We suggest that at least in some instances, these features simply move out-of-view of the narrow transmission filter, which is supported by the lateral motion observed in Fig.~\ref{eg2}. Since, the observed lateral shift could be due to LOS effects where the features are moving towards and away from us and due to our viewing angle we see a motion which could be interpreted as a kink wave, e.g. \citet{2014ApJ...788....9G}.

\subsection{Possible association with waves - further justifying flux tube behaviour}

A possible explanation for the observed lateral motion could be the presence of transverse kink waves \citep{Kukhianidze2006,2007A&A...474..627Z}. \citet{2008ASPC..397...27S} looked at the lateral motion of spicules as a function of height above the limb and found that such motion became more prominent with height. Furthermore, they noted a period of oscillation between 1 to 4 min, an amplitude of about 1$\arcsec$\, and a maximum lateral velocity of $\sim$15 -- 25 km s$^{-1}$, very similar to the values derived in this work. \citet{1971SoPh...18..403N} showed that some spicules change their position along the slit, thereby showing a real horizontal component of velocity. Furthermore, they concluded that spicules can execute quasi-periodic oscillations parallel to the limb with a period of $\sim$1 min, an amplitude of 1\arcsec\ and velocities around 10 -- 15 km s$^{-1}$, similar to the more recent work
cited above. In our observations we see that the amplitude of lateral motion is around 230 km (maximum displacement from the mean position observed in CRISPEX) with a period of 20s and velocities between 15 -- 30 km s$^{-1}$. \citet{Kukhianidze2006} looked at spicules at 8 different heights, from 3800 -- 8700 km. They found that $\sim$20\% of the measured height series showed a periodic spatial distribution in the Doppler velocities with a wavelength of 3500 km and periods in the range of 35 -- 70 s, suggesting transverse kink waves. Later work by \citet{2012ApJ...750...51K, 2013ApJ...779...82K,2012NatCo...3E1315M} reported propagating and standing transverse oscillations in H$\alpha$ core mottles, which are believed to be the disk counterparts of classical (Type--I) limb spicules.  
They interpreted the observed transverse motions in terms of propagating/standing kink waves and
used the measured wave properties to estimate chromospheric plasma parameters by employing the solar magneto--seismology method.

\citet{2007Sci...318.1574D} suggested that Type--II spicules display signatures of both torsional and swaying motions. The spicules in Fig.~\ref{case23_wl} and Fig.~\ref{case1_wl}, are some of the few showing simultaneous or alternate appearances in the red and blue wing. The one in Fig.~\ref{case23_wl} appears simultaneously at $\pm$ 774 m\AA\, in panel at 07:57:20.638 UT. However, it disappears in the red wing in panel at 07:57:20.638 UT but continues to appears from 07:57:24.618 UT onwards in the blue wing. This behaviour is interpreted as an example of torsional motions by \citet{2007Sci...318.1574D}. For example in Fig.~4 of their paper a spicule is seen changing its sign from being blue-shifted to red-shifted at $\pm$55 km s$^{-1}$ showing the presence of swaying motions of the order 15 -- 20 km s$^{-1}$, and torsional motion of order 25 -- 30 km s$^{-1}$. \citet{2007Sci...318.1574D} found the period of such waves to be of the order of 100 s.

\section{Summary and Conclusions}\label{Summ}

The data shows that highly-dynamic spicules appear at different locations within the H$\alpha$ line profile within a few seconds, sometimes  with a temporal offset. They have very short lifetimes of up to 20s and length/height around 3500 km, which could imply that these are in some cases a subset of RBEs. In some instances, these features show a time delay in their appearance in the blue and red wings of the H$\alpha$ line by 3--5s, while some features show simultaneous or/and alternate appearances in the red and blue wing. This would indicate some sort of wave--like behaviour. In some instances, the features seem to re-emerge at the same location several tens of seconds later. We suggest that these observations can be interpreted as the movement of highly dynamic spicules moving in and out of the narrow 60 m\AA\, transmission filter used to observe in different parts of the H$\alpha$ line profile. The LOS velocity component of the observed fast chromospheric features, manifested as Doppler shifts, are responsible for their appearance in the red and blue wings of H$\alpha$  line.

In a companion paper we address in more detail the apparent wave activity. We were unable to use the present Ca~{\sc ii} dataset as it did not extend sufficiently far into the wings, however, new studies taken at other wavelengths (corresponding to different formation heights) of Ca~{\sc ii} or with Interface Region Imaging Spectrograph (IRIS) in say Si~{\sc iv} or Mg~{\sc ii} could provide valuable information on whether the above explanation is likely. 

\begin{acknowledgements}  
Armagh Observatory is grant-aided by the N. Ireland Department of Culture, Arts and Leisure. The Swedish 1-m Solar Telescope is operated on the island of La Palma by the Institute for Solar Physics of Stockholm University in the Spanish Observatorio del Roque de los Muchachos of the Instituto de Astrofsica de Canarias. The authors wish to acknowledge the DJEI/DES/SFI/HEA Irish Centre for High-End Computing (ICHEC) for the provision of computing facilities and support. We also like to thank STFC for a studentship and PATT T\&S and the Solarnet project which is supported by the European Commission's FP7 Capacities Programme under Grant Agreement number 312495 for T\&S. JS is funded by the Leverhulme Trust. D.K. would like to acknowledge funding from the European Commission’s Seventh Framework Programme (FP7/2007- 2013) under grant agreement No. 606862 (F-CHROMA). JS would like to thank Dr. Pit S\"utterlin for valuable advice. We would like to thank the referee for valuable inputs into the paper. \end{acknowledgements}

\bibliographystyle{aa}
\bibliography{references}
\begin{appendix}

\section{Description of the events}

\begin{table*} 
 \begin{sideways}
\begin{minipage}{26 cm}
\caption{Occurrence time (hh:mm:ss.s UT) chart of the events observed on 2014 June 10. The duration or lifetime (s) at each position is given in the brackets. \label{A1}}
\begin{tabular}{||l|| c c c c c c ||l ||l||}
\hline\hline
Case no & & &    & m\AA\ from H $\alpha$ line core &  & & Position &  Comment \\

  &  -1032m \AA &-774m \AA & -516m\AA & +516m \AA & + 774 m\AA & + 1032m\AA  &  (nx",ny") &  \\ \hline
1 & 08:07:23 (11.1) & 08:07:19 (14.8) & 08:07:23 (3.9) & 08:07:27
(3.9)& 08:07:34 (14.8) & & 29,22 & Isolated Spicule.  red \& blue-wing \\
&  & & & & & & & signature separated by one pixel and 11.1 s.\\

\hline
2 & & 07:30:29 (18.5) & 07:30:21 (3.9) & 07:30:13 (18.5)& 7:30:37 (3.9) & 	 & 24,16  & Co-spatial red \& blue wing signatures. \\
\hline
3 & 07:30:53 (14.8) & 07:30:49 (25.9) &&&&& 6,24 &Faint and recurring event   \\
&  & & & & & & & (reoccurs after 30 s). \\
\hline
4 & & 07:35:54 (11.1) & 07:35:54 (11.1) && 07:36:02 (11.1) & &11,21 & Reappears after 37 s one pixel away.   \\
&  & & & & & & & -1023 m\AA\, signature is smaller in size. \\
\hline
5 & 07:35:18 (7.4) & 07:35:14 (14.8) &&&&& 20,22 & -1023 m\AA\, signature is smaller in size. \\
&  & & & & & & & Red wing shows two parallel features\\
&  & & & & & & & at 07:35:06 UT, \\
&  & & & & & & & one pixel away on both sides. \\
\hline
6 && 07:29:25.456 (22.2) &&&&& 25,20  & Observed only in speckled data.\\
\hline
7 & & 07:43:52.740 (18.5) &&&&& 12,31 &  Observed only in speckled data. \\
\hline
8 & 08:06:46 (7.4) & 08:06:46(22.2) & 08:07:07 (7.4) & 08:06:46 (22.2) & 08:06:46 (22.2) & 08:06:59 (22.2) & 35,18&Stronger signature in red wing.\\
&  & & & & & & & Group of fast features \\
\hline
9 & & 07:54:22 (7.4) & 07:54:26 (11.1) & 07:54:50 (11.1) & 07:54:42 (11.1) & & 24,9& Recurring feature. Red wing one pixel away. \\ 
&  & & & & & & & -1023 m\AA\, signature is smaller in size. \\
\hline
10 & & 08:04:04 (22.2) & &&& & 24,12 & Fast feature near a flow.\\
\hline
11 & 07:55:46 (3.9) & 07:55:46 (18.5) & 07:55:42 (22.2) && 07:55:46 (3.9) && 13,33 & Isolated feature with lateral movement \\ 
&  & & & & & & &  Recurring structure \\
 &  & & & & & & & at 07:56:09 UT for 3.9s. \\
&  & & & & & & &  Smaller signature at +777 m\AA\,. \\

\hline
12 & 07:54:50 (3.9) & 07:54:46 (18.5) & &07:54:38 (14.8) & 07:54:50 (11.1) &&25,10 & Lateral movement \\
\hline
13 & 08:02:18 (3.9) & 08:02:14 (11.1) & 08:02:14 (3.9) && 08:02:14 (11.1)& &39,53 & Very faint red wing signature. Flow?  \\
\hline
14 & & 08:03:05 (11.1) &&&&&21,20&Curved event. \\
\hline
15 && 07:56:53 (3.9) &&& 07:56:57 (7.4) & 07:56:57 (7.4 ) & 6,25 &Small feature. Very fast appearance \\
&  & & & & & & & and disappearance. \\
&  & & & & & & & (Blue wing signature is smaller in size)\\
\hline
16 &&&07:52:40 (3.9) & 07:52:36 (7.4)& 07:52:36 (7.4)& 07:52:36 (3.9) &43,42&Small, thin event. one pixel shift at 516 m\AA.\\
\hline
17 && &&&07:58:40 (14.8)&07:58:44 (7.4)&21,23 & Long and thin event. \\
\hline
18 && 07:56:57 (18.5)&&&07:57:01 (18.5)&&36,24 & Red \& blue-wing signatures offset\\
&  & && & & & &  by two pixels. Flow-like in blue wing. \\
\hline
19 &&&&& 07:55:50 (11.1)& 07:55:54 (11.1) & 36,24 & Red \& blue wing signatures \\
&  & & & & & & & are different.  \\
\hline
20 & 07:57:29 (3.9) &07:57:25 (11.1) & & 07:57:13  & 07:57:17 (7.4) & 07:57:33 (7.4)  &45,35& Slight lateral movement. -1023 m\AA\, \\
&  & & & & & & & signature is smaller in size than others.  \\
\hline
21 && 08:04:00 (7.4) &&& 08:02:06 (7.4) & 08:02:06 (3.9) &29,13 & Amazing shift in red wing. \\
&  & & & & & & & Blue wing shifted by one pixel.  \\
\hline
22 && 07:35:54 (7.4) & 07:35:54 (14.8)&& 07:36:02 (14.8)& &12,22 &Same as case 1 but only in blue wing.\\
\hline
23 & & 08:02:37 (3.9) &&&&&43,45 & Fast feature near another fast feature.\\
&  & & & & & & & (case 12)  \\
\hline

\end{tabular} 
\end{minipage}
\end{sideways}
\end{table*}

\begin{table*}
\begin{sideways}
\begin{minipage}{27 cm} 
\centering 
\caption{Occurrence time (hh:mm:ss.s UT) chart of the events observed on 2014 June 7. The duration or lifetime (s) at each position is given in the brackets.  \label{A2}}
\begin{tabular}{||l|| c c c c c c ||l ||l||}
\hline\hline
Case no & & &    & m\AA\ from H $\alpha$ line core &  & & Position &  Comment \\

 &  -1032 m\AA &-774 m\AA & -516 m\AA & +516 m\AA & + 774 m\AA & + 1032 m\AA  &  (nx",ny") &  \\ \hline
 
1 & 12:24:06 (30)	& 12:24:11 (40) & 12:23:56 (35) &  & & & 39,15 &Isolated feature. Both sides   \\
&  & & & & & & & show lateral movement.  Looks like\\
&  & & & & & & & re-emergence with 10s gap in --774 m\AA.  \\
\hline
2 & 12:01:20 (25) & 12:01:15 (40) & &&&&6,6& Feature is near another feature. \\ 
&  & & & & & & & Size of feature is different at the different \\
&  & & & & & & & wavelengths. Shows lateral movement. \\
\hline	
3 & 12:06:44 (10) & 12:06:39 (15) &&&&&9,29& Straight and isolated feature.\\
\hline
4 & 12:21:24 (20) & 12:21:14 (30) &&&&&46,4& This is horizontal feature  \\
 &  &  & & & & & & with a very nice lateral motion.\\
 &  &  & & & & & & (looks like swaying motion)\\
\hline
5 &12:10:31 (15) & 12:10:26 (30) &&&&&11,26& Nice clear feature but\\
 &  &  & & & & &  & close to other relatively slow spicules.\\
 &  &  & & & & &  & Possible lateral motion.\\
  &  &  & & & & &  & -1032 m\AA\ signature is smaller. \\
\hline
6 && 12:04:12 (35) &&&12:04:32 (5) &&12,14 &A Very strong lateral movement \\
&  &  & & & & & & Very faint signature in red-wing.\\
\hline
7 && & & &12:20:14 (15) &&23,20& Lateral shift of one pixel. \\
\hline
8 & 12:31:37 (10)  & 12:31:17 1(25) &&& & &24,19 &Only blue wing signature  \\
\hline
9 && 12:34:19 (25) &&&&&30,16& Only blue wing feature  \\
&  &  & & & & & & near a flow-like structure.\\
\hline
10 & 12:16:46.142(25) & 12:16:46.142(25) &&& 12:16:51.217(15) & 12:16:56.266(5) &29.5,& Very nice red-blue wing   \\
&  & & & & & & & signature. Next to slow jets.\\
\hline
11 &&12:08:05 (20) &&&&&27,27&A feature near other slow moving spicules. \\
&  & & & & & & & Shows repeats after 60 s for 5 s.\\
\hline
12 &&&&&11:58:48 (15) & 11:58:53 (10) &31,9&Feature is seen very clearly in the red wing \\
&  & & & & & & & One pixel lateral shift.\\
\hline
13 &&&& & 12:05:28 (25) & 12:05:38 (5) &25,28& Curved feature in red wing. \\
\hline
14 & & && & 11:59:18 (20) & 11:59:23 (15) &26,19&Very long and slender feature \\
  & & & & & & & & which shows lateral movement in red wing.\\
  & & & & & & & & Blue-wing signature is shifted by one pixel\\
\hline
15 &&&&12:11:12 (15) & 12:11:17 (20) & 12:11:22 (10) & 31,18& Long feature which \\
& & & & & & & &shows lateral movement. Repeats after 10 s. \\
\hline
16 &12:00:14 (15) &12:00:14 (20) &&&12:00:19 (15) &&27,19&Shows a repeat after 105 s one pixel away\\
\hline
17 && 11:59:53 (25) &&&&&10,26 & Repeating features (repeats after 20 s)  \\ 
\hline
\end{tabular} 
\end{minipage}
\end{sideways}
\end{table*}

\end{appendix}

\end{document}